\documentclass[12pt,onecolumn,draftclsnofoot]{IEEEtran}

\usepackage{times}
\usepackage{amsmath, amsthm}
\usepackage{amsfonts}
\usepackage{mathrsfs}
\usepackage{latexsym}
\usepackage{amssymb}
\usepackage{enumerate}
\usepackage{upref}
\usepackage{graphicx}
\usepackage{color}
\usepackage{scrextend}
\usepackage{algorithm}
\usepackage{algpseudocode}
\usepackage{verbatim}

\usepackage{textcomp}
\usepackage[paper=letterpaper,top=1.1in,bottom=0.8in,left=0.8in,right=0.8in]{geometry}

\theoremstyle{definition}

\newtheorem{thm}{Theorem}

\newtheorem{lem}{Lemma}

\newtheorem{defn}{Definition}
\newtheorem{exa}{Example}

\newtheorem{rem}{Remark}

\newcommand\bl[1]{{\color{blue}#1}}

\begin{document}

  \title{On the Sub-Packetization Size and the Repair Bandwidth of  Reed-Solomon Codes}
  \author{ 
  \IEEEauthorblockN{Weiqi Li, Zhiying Wang, Hamid Jafarkhani}\\
\IEEEauthorblockA{Center for Pervasive Communications and Computing (CPCC)  \\ University of California, Irvine, USA
\\ \{weiqil4, zhiying, hamidj\}@uci.edu
} 
}

\maketitle

\begin{abstract}
Reed-Solomon (RS) codes are widely used in distributed storage systems. In this paper, we study the repair bandwidth and sub-packetization size of RS codes. The repair bandwidth is defined as the amount of transmitted information from surviving nodes to a failed node. The RS code can be viewed as a polynomial over a finite field $GF(q^\ell)$ evaluated at a set of points, where $\ell$ is called the sub-packetization size. Smaller bandwidth reduces the network traffic in distributed storage, and smaller $\ell$ facilitates the implementation of RS codes with lower complexity. Recently, Guruswami and Wootters proposed a repair method for RS codes when the evaluation points are the entire finite field. While the sub-packization size can be arbitrarily small, the repair bandwidth is higher than the minimum storage regenerating (MSR) bound.
Tamo, Ye and Barg achieved the MSR bound but the sub-packetization size grows faster than the exponential function of the number of the evaluation points. In this work, we present code constructions and repair schemes that extend these results to accommodate different sizes of the evaluation points. In other words, we design schemes that provide points in between. These schemes provide a flexible tradeoff between the sub-packetization size and the repair bandwidth. In addition, we generalize our schemes to manage multiple failures.

\end{abstract}

\section{Introduction}

Erasure codes are ubiquitous in distributed storage systems because they can efficiently store data while protecting against failures. Reed-Solomon (RS) code is one of the most commonly used codes because it achieves the Singleton bound  \cite{macwilliams1977theory} and has efficient encoding and decoding methods, see, e.g.,  \cite{ryan2009channel, guruswami1998improved}. Codes matching the Singleton bound are called maximum distance separable (MDS) codes, and they have the highest possible failure-correction capability for a given redundancy level. 
 In distributed storage, every code word symbol corresponds to a storage node, and communication costs between storage nodes need to be considered when node failures are repaired. In this paper, we study the \emph{repair bandwidth} of RS codes, defined as  the amount of transmission required to repair a single node erasure, or failure, from all the remaining nodes (called helper nodes). 

For a given erasure code, when each node corresponds to a single finite field symbol over $\mathbb{F}=GF(q^\ell)$, we say the code is scalar; when each node is a vector of finite field symbols in $\mathbb{B}=GF(q)$ of length $\ell$, it is called a vector code or an array code. In both cases, we say the \emph{sub-packetization size} of the code is $\ell$. Here $q$ is a power of a prime number.
Shanmugam et al. \cite{shanmugam2014repair} considered the repair of scalar codes for the first time.
Recently, Guruswami and Wootters  \cite{guruswami2017repairing} proposed a repair scheme  for RS codes. The key idea of both papers is that: rather than directly using the helper nodes as symbols over $\mathbb{F}$ to repair the failed node, one treats them as vectors over the subfield $\mathbb{B}$. 
Thus, a helper may transmit less than $\ell$ symbols over $\mathbb{B}$, resulting in a reduced bandwidth.
For an RS code with length $n$ and dimension $k$ over the field $\mathbb{F}$, denoted by $RS(n,k)$, \cite{guruswami2017repairing} achieves the repair bandwidth of $n-1$ symbols over $\mathbb{B}$.
Moreover, when $n=q^\ell$ (called the full-length RS code) and $n-k=q^{\ell-1}$, 
the scheme provides the optimal repair bandwidth. Dau and Milenkovic \cite{dau2017optimal} improved the scheme 
such that the repair bandwidth is optimal for the full-length RS code and any $n-k=q^s, 1\leq s \leq \log_q(n-k)$. 

For the full-length RS code, the schemes in \cite{guruswami2017repairing} and \cite{dau2017optimal} are optimal for single erasure. 
However, the repair bandwidth of these schemes still has a big gap from the minimum storage regenerating (MSR) bound derived in \cite{dimakis2010network}. 
In particular, for an arbitrary MDS code, the repair bandwidth $b$, measured in the number of symbols over $GF(q)$, is lower bounded by
\begin{align}\label{MSR bound}
b\geq \frac{\ell(n-1)}{n-k}.
\end{align}  
An MDS code satisfying the above bound is called an MSR code.
In fact, most known MSR codes are vector codes, see \cite{rashmi2011optimal,papailiopoulos2013repair,Zigzag_Codes_IT,wang2014explicit,rawat2016progress,goparaju2017minimum, ye2016nearly}.
For the repair of RS codes, Ye and Barg proposed a scheme that asymptotically approaches the MSR bound as $n$ grows \cite{ye2016explicit} when the sub-packetization size is $\ell=(n-k)^n$. 
Tamo et al. \cite{tamo2017optimal} provided an RS code repair scheme achieving the MSR bound when the sub-packetization size is $\ell \approx n^n$. 

The repair problem for RS codes can also be generalized to multiple erasures. In this case, the schemes in \cite{dau2018repairing} and \cite{mardia2018repairing} work for the full-length code, \cite{zorgui2019centralized} and \cite{zorgui2018achievability} work for centralized repair, and \cite{ye2017repairing} proposed a scheme achieving the multiple-erasure MSR bound. 


\textbf{Motivation}:
A flexible tradeoff between the sub-packetization size and the repair bandwidth is an open problem: Only the full-length RS code with high repair bandwidth and the MSR-achieving RS code with large sub-packetization are established. Our paper aims to provide more points between the two extremes -- the full-length code and the MSR code. One straightforward method is to apply the schemes of \cite{guruswami2017repairing} and \cite{dau2017optimal} to the case of $\ell > \log_q n$ with fixed $(n,k)$. However, the resulting normalized repair bandwidth $\frac{b}{\ell (n-1)}$ grows with $\ell$, contradictory to our intuition that larger $\ell$ implies smaller normalized bandwidth.


The need for small repair bandwidth is motivated by reducing the network traffic in distributed storage \cite{dimakis2010network}, and the need for the 
small sub-packetization is due to the complexity in field arithmetic operations, discussed below. It is demonstrated that the time complexity of multiplications in larger fields are much higher than that of smaller fields \cite{gashkov2013complexity}.
Moreover, multiplication in Galois fields are usually done by pre-computed look-up tables and the growing field size has a significant impact on the space complexity of multiplication operations.  Larger fields require huge memories for the look-up table. For example, in $GF(2^{16})$, $8$ GB are required for the complete table, which is impractical in most current systems \cite{greenan2008optimizing}. Some logarithm tables and sub-tables are used to alleviate the memory problems for large fields, while increasing the time complexity at the same time \cite{greenan2008optimizing},  \cite{plank2013screaming}, \cite{luo2012efficient}. For example, in the Intel SIMD methods, multiplications over $GF(2^{16})$ need twice the amount of operations as over $GF(2^8)$, and multiplications over $GF(2^{32})$ need $4$ times the amount of operations compared to $GF(2^8)$, which causes the multiplication speed to drop significantly when the field size grows \cite{plank2013screaming}. 

To illustrate the impact of the sub-packetization size on the complexity, let us take encoding for example. To encode a single parity check node, we need to do $k$ multiplications and $k$ additions over $GF(q^\ell)$. For a given systematic $RS(n,k)$ code over $GF(q^\ell)$, we can encode $k\ell \log_2q$ bits of information by multiplications of $(n-k)k\ell \log_2q$ bits and additions of $(n-k)k\ell \log_2q$ bits. So, when $M$ bits are encoded into $RS(n,k)$ codes, we need $M/(k\ell \log_2q)$ copies of the code and we need multiplications of $M(n-k)$ bits and additions of $M(n-k)$ bits in $GF(q^\ell)$ in total. Although the total amount of bits we need to multiply is independent of $\ell$, the complexity over a larger field is higher in both time and space.
For a simulation of the RS code speed using different field sizes on different platforms, we refer the readers to \cite{plank2009performance}. The results suggest that RS codes have faster implementation in both encoding and decoding for smaller fields.

Besides the complexity, the small sub-packetization level also has many advantages such as easy system implementation, great flexibility and bandwidth-efficient access to missing small files \cite{guruswami2017mds}, \cite{khan2012rethinking}, which makes it important in distributed storage applications. 

As can be seen from the two extremes, a small sub-packetization level also means higher costs in repair bandwidth, and not many other codes are known besides the extremes. For vector codes, Guruswami,  Rawat \cite{guruswami2017mds} and Li, Tang \cite{li2019systematic} provided small sub-packetization codes with small repair bandwidth, but only for single erasure.  Kralevska et al. \cite{kralevska2018hashtag} also presented a tradeoff between the sub-packetization level and the repair bandwidth for the proposed HashTag codes implemented in Hadoop. For scalar codes, Chowdhury and Vardy \cite{chowdhury2017improved} extended Ye and Barg's MSR scheme \cite{ye2016explicit} to a smaller sub-packetization size, but it only works for certain redundancy $r$ and single erasure.

\textbf{Contributions}:
In this work, we first design three single-erasure RS repair schemes, using the cosets of the multiplicative group of the finite field $\mathbb{F}$. Note that the RS code can be viewed as $n$ evaluations of a polynomial over $\mathbb{F}$. The evaluation points of the three schemes are part of one coset, of two cosets, and of multiple cosets, respectively, so that the evaluation point size can vary from a very small number to the whole field size. In the schemes designed in this paper, we have a parameter $a$ that can be tuned, and provides a tradeoff between the sub-packetization size and the repair bandwidth.
\begin{itemize}
\item For an $RS(n,k)$ code, our first scheme achieves the repair bandwidth $\frac{\ell}{a}(n-1)(a-s)$ for some $a, s$ such that $n<q^a, r \triangleq n-k>q^s$ and  $a$ divides $\ell$. Specifically, for the $RS(14,10)$ code used in Facebook \cite{sathiamoorthy2013xoring}, we achieve repair bandwidth of $52$ bits with $\ell=8$, which is $35\%$ better than the naive repair scheme.
\item Our second scheme reaches the repair bandwidth of $(n-1)\frac{\ell+a}{2}$ for some $a$ such that $n\leq2(q^a-1)$, $a$ divides $\ell$ and $\frac{\ell}{r}<a$. 

\item The first realization of our third scheme attains the repair bandwidth of $\frac{\ell}{r}(n+1+(r-1)(q^a-2))$ when $n\leq(q^a-1)\log_{r}\frac{\ell}{a}$. Another realization of the third scheme attains the repair bandwidth of $\frac{\ell}{r}(n-1+(r-1)(q^a-2))$ where $\ell\approx a   (\frac{n}{q^a-1})^{ (\frac{n}{q^a-1})}$. The second realization can also be generalized to any $d$ helpers, for $k \le d \le n-1$.
\end{itemize}
We provide characterizations of linear multiple-erasure repair schemes, and propose two schemes for multiple erasures, where the evaluation points are in one coset and in multiple cosets, respectively. Again, the parameter $a$ is tunable.
\begin{itemize}
\item We prove that any linear repair scheme for multiple erasures in a scalar MDS code is equivalent to finding a set of dual codewords satisfying certain rank constraints.
\item For an $RS(n,k)$ code with $e<\frac{1}{a-s}\sqrt{\log_q n}$ erasures, our first scheme achieves the repair bandwidth $\frac{e\ell}{a}(n-e)(a-s)$ for some $a, s$ such that $n<q^a, r = n-k>q^s$ and  $a$ divides $\ell$. 
\item For an $RS(n,k)$ code, our second scheme works for $e\leq n-k$ erasures and $n-e$ helpers. The repair bandwidth depends on the location of the erasures and in most cases, we achieve $\frac{e\ell}{d-k+e}(n-e+(n-k+e)(q^a-2))$ where $\ell\approx a  (\frac{n}{q^a-1})^{ (\frac{n}{q^a-1})}$ and $a$ divides $\ell$.
\item We demonstrate that repairing multiple erasures simultaneously is advantageous compared to repairing single erasures separately.
\end{itemize}
The comparison of our schemes, as well as the comparison to previous works, are shown in Tables \ref{table of comparision1} and \ref{table of comparison for multiple erasures}, and are discussed in more details in Sections \ref{sec:single_numerical} and \ref{sec:multiple_numerical}. 

\begin{table*}
\newcommand{\tabincell}[2]{\begin{tabular}{@{}#1@{}}#2\end{tabular}}
\centering
\caption{\normalfont  Comparison of different schemes for single erasure. When $a=\ell$, our scheme in one coset is the scheme in \cite{guruswami2017repairing}, \cite{dau2017optimal}. When $a=1$, our schemes in multiple cosets is the schemes in \cite{ye2016explicit}, \cite{tamo2017optimal}. }\label{table of comparision1}
\begin{tabular}{|c|c|c|c|}

\hline

&repair bandwidth& code length & restrictions \\
\hline
\tabincell{c} {Schemes in \cite{guruswami2017repairing}, \cite{dau2017optimal}}& $(n-1)(\ell-s)$  & $n\leq q^\ell$ &$q^s\leq r$   \\
\hline
\tabincell{c} {Scheme in \cite{ye2016explicit}}& $<\frac{\ell}{r}(n+1)$  & $n=\log_r\ell$&   \\
\hline
\tabincell{c} {Scheme in \cite{tamo2017optimal}}& $\frac{\ell}{r}(n-1)$  & $n^n\approx\ell$&  \\
\hline
\tabincell{c} {Our scheme in \\one coset}& $\leq \frac{\ell}{a}(n-1)(a-s)$  & $n\leq(q^a-1)$& $q^s\leq r, a|\ell$  \\
\hline
\tabincell{c}{Our scheme in \\two cosets} & $<(n-1)\frac{\ell+a}{2}$ & $n\leq2(q^a-1)$& $\frac{\ell}{r}\leq a ,a|\ell$ \\
\hline
\tabincell{c}{Our scheme in \\multiple cosets 1}& $ <\frac{\ell}{r}(n+1+(r-1)(q^a-2))$ & \tabincell{c}{$n\leq(q^a-1)m$ } & \tabincell{c}{$\ell/a=r^m$ \\for some integer $m$}  \\
\hline
\tabincell{c}{Our scheme in \\multiple cosets 2}& $ \frac{\ell}{r}(n-1+(r-1)(q^a-2))$ & \tabincell{c}{$n\leq(q^a-1)m$ } & \tabincell{c}{$\ell/a\approx m^m$ \\for some integer $m$}  \\
\hline
\end{tabular}

\end{table*}

\begin{table*}
\newcommand{\tabincell}[2]{\begin{tabular}{@{}#1@{}}#2\end{tabular}}
\centering
\caption{\normalfont  Comparison of different schemes for multiple erasures. When $a=\ell$ and $s=\ell$ our scheme in one coset is the scheme 1 in \cite{mardia2018repairing}. When $a=1$, our schemes in multiple cosets is the scheme in \cite{ye2017repairing}. }\label{table of comparison for multiple erasures}
\begin{tabular}{|c|c|c|c|}

\hline

&repair bandwidth& code length & restrictions\\
\hline
\tabincell{c} {Scheme 1 in \cite{mardia2018repairing}}& $\le(n-e)e-\frac{e(e-1)(q-1)}{2}$  & $n\le q^\ell$ &$q^{\ell-1}\leq r, e<\sqrt{\log_q n}$   \\
\hline
\tabincell{c} {Scheme 2 in \cite{mardia2018repairing}}& $\le\min \limits_{e'\ge e}((n-e')(\ell- \lfloor \log_q(\frac{n-k+e'-1}{2e'-1}) \rfloor))$  & $n\leq q^\ell$ &   \\
\hline
\tabincell{c} {Scheme in \cite{ye2017repairing}}& $\frac{ed\ell}{d-k+e}$  & $n^n\approx\ell$&  \\
\hline
\tabincell{c} {Our scheme for multiple erasures \\in one coset}& $\leq \frac{e\ell}{a}(n-e)(a-s)$  & $n\leq(q^a-1)$& $q^s\leq r, a|\ell,e<\frac{1}{a-s}\sqrt{\log_q n}$ \\
\hline
\tabincell{c}{Our scheme for multiple erasures \\in multiple cosets}& $\frac{e\ell}{n-k}(n-e+(n-k+e)(q^a-2))$ & \tabincell{c}{$n\leq(q^a-1)m$ } & \tabincell{c}{$\ell/a\approx m^m$ \\for some integer $m$}  \\
\hline
\end{tabular}

\end{table*}

The paper is organized as follows. In Section II, we briefly review the linear repair of RS codes and provide the preliminaries used in this paper. In Section III, we present three RS repair schemes for single erasure. Then, we discuss the repair schemes for multiple erasures in Section IV. In Section V, we provide the conclusion.

{\bf Notation}: Throughout this paper, for positive integer $i$, we use $[i]$ to denote the set $\{1,2,\dots,i\}$. 
For integers $a,b$, we use $a \mid b$ to denote that $a$ divides $b$. {For real numbers $a_n,b_n$, which are functions of $n$, we use $a\approx b$ to denote $\lim_{n \to \infty} \frac{a_n}{b_n}=1$.}
For sets $A \subseteq B$, we use $B/A$ to denote the difference of $A$ from $B$. For a finite field $\mathbb{F}$, we denote by $\mathbb{F}^*=\mathbb{F}/\{0\}$ the corresponding multiplicative group. We write $\mathbb{E} \le \mathbb{F}$ for $\mathbb{E}$ being a subfield of $\mathbb{F}$.
For element $\beta \in\mathbb{F}$ and $E$ as a subset of $\mathbb{F}$, we denote $\beta E=\{\beta s, \forall s\in E\}$. $A^T$ denotes the transpose of the matrix $A$.

\section{Preliminaries}
In this section, we review the linear repair scheme of RS code in \cite{guruswami2017repairing}, and provide a basic lemma used in our proposed schemes.

The \emph{Reed-Solomon code} $RS(A,k)$ over $\mathbb{F}=GF(q^\ell)$ of dimension $k$ with $n$ evaluation points $A=\{\alpha_{1},\alpha_{2},\dots,\alpha_{n}\}\subseteq \mathbb{F}$ is defined as
\begin{align*}
RS(A,k)=\{ & (f(\alpha_{1}),f(\alpha_{2}),\dots,f(\alpha_{n})): f\in \mathbb{F}[x], \deg(f)\le k-1 \},
\end{align*}
where $\deg()$ denotes the degree of a polynomial, $f(x)=u_{0}+u_{1}x+u_{2}x^2+\dots+u_{k-1}x^{k-1}$, and $u_{i} \in \mathbb{F}, i=0,1, \dots,k-1$ are the messages. Every evaluation symbol $f(\alpha), \alpha \in A,$ is called a code word symbol or a storage node. 
The \emph{sub-packetization size} is defined as $\ell$, and $r \triangleq n-k$ denotes the number of parity symbols.

Assume $e$ nodes fail, $e \le n-k$, and we want to recover them. The number of helper nodes are denoted by $d$.
The amount of information transmitted from the helper nodes is defined as the \emph{repair bandwidth} $b$, measured in the number of symbols over $GF(q)$. All the remaining $n-e=d$ nodes are assumed to be the helper nodes unless stated otherwise. We define the \emph{normalized repair bandwidth} as $\frac{b}{\ell d}$, which is the average fraction of information transmitted from each helper. By \cite{dimakis2010network, cadambe2013asymptotic}, the minimum storage regenerating (MSR) bound for the bandwidth is
\begin{align}
b \ge \frac{e \ell d}{d-k+e}.
\end{align}
As mentioned before, codes achieving the MSR bound require large sub-packetization sizes.
In this section, we focus on the single erasure case.

Assume $\mathbb{B} \le \mathbb{F}$, namely, $\mathbb{B}$ is a subfield of $\mathbb{F}$. A linear repair scheme requires some symbols of the subfield $\mathbb{B}$ to be transmitted from each helper node \cite{guruswami2017repairing}. If the symbols from the same helper node are linearly dependent, the repair bandwidth decreases. 
In particular, the scheme uses dual code to compute the failed node and uses trace function to obtain the transmitted subfield symbols, as detailed below.

Assume $f(\alpha^\ast)$ fails for some $\alpha^\ast \in A$.
For any polynomial $p(x) \in \mathbb{F}[x]$ of which the degree is smaller than $r$, $(\upsilon_{1}p(\alpha_{1}),\upsilon_{2}p(\alpha_{2}),\dots,\upsilon_{n}p(\alpha_{n}))$ is a dual codeword of $RS(A,k)$, where $\upsilon_{i}, i\in[n]$ are non-zero constants determined by the set $A$ (see for example \cite[Thm. 4 in Ch.10]{macwilliams1977theory}). We can thus repair the failed node $f(\alpha^\ast)$ from 
\begin{align}\label{eq_dual}
\upsilon_{\alpha^{\ast}}p(\alpha^{\ast})f(\alpha^{\ast})=-\sum\limits_{i=1,\alpha_i\neq\alpha^{\ast}}^{n}\upsilon_{i}p(\alpha_{i})f(\alpha_{i})
\end{align}
The summation on the right side means that we add all the $i$ elements from $i=1$ to $i=n$ except when $\alpha_i\neq\alpha^{\ast}$.

The trace function from $\mathbb{F}$ onto $\mathbb{B}$ is defined as
\begin{equation}\label{trace definition}
tr_{\mathbb{F}/\mathbb{B}}(\beta)=\beta+\beta^q+\dots+\beta^{q^{\ell-1}},
\end{equation}
where $\beta \in \mathbb{F}$, $\mathbb{B}=GF(q)$ is called the \emph{base field}, and $q$ is a power of a prime number. It is a linear mapping from  $\mathbb{F}$ to $\mathbb{B}$ and satisfies
\begin{equation}
tr_{\mathbb{F}/\mathbb{B}}(\alpha \beta)=\alpha tr_{\mathbb{F}/\mathbb{B}}(\beta)
\end{equation}
for all $\alpha \in \mathbb{B}$.

We define the rank $rank_{\mathbb{B}}(\{\gamma_1,\gamma_2,...,\gamma_i\})$ to be the cardinality of a maximal subset of $\{\gamma_1,\gamma_2,...,\gamma_i\}$ that is linearly independent over $\mathbb{B}$. For example, for $\mathbb{B}=GF(2)$ and $\alpha \notin \mathbb{B}$, $rank_{\mathbb{B}}(\{1,\alpha,1+\alpha\})=2$ because the subset $\{1, \alpha\}$ is the maximal subset that  is linearly independent over $\mathbb{B}$ and the cardinality of the subset is $2$.

Assume we use polynomials $p_j(x)$, $j\in[\ell]$ to generate $\ell$ different dual codewords, called \emph{repair polynomials}. Combining the trace function and the dual code, we have
\begin{align}\label{trace equation}
&tr_{\mathbb{F}/\mathbb{B}}(\upsilon_{\alpha^{\ast}}p_{j}(\alpha^{\ast})f(\alpha^{\ast}))=-\sum\limits_{i=1,\alpha_i\neq\alpha^{\ast}}^{n}tr_{\mathbb{F}/\mathbb{B}}(\upsilon_{i}p_{j}(\alpha_{i})f(\alpha_{i})).
\end{align}
In a repair scheme, the helper $f(\alpha_i)$ transmits 
\begin{align} \label{eq:helper trans}
\{tr_{\mathbb{F}/\mathbb{B}}(\upsilon_{i}p_{j}(\alpha_{i})f(\alpha_{i})):j\in[\ell]\}.    
\end{align}
Suppose $\{\upsilon_{\alpha^{\ast}}p_{1}(\alpha^\ast),$ $\upsilon_{\alpha^{\ast}}p_{2}(\alpha^\ast),$ $\dots,\upsilon_{\alpha^{\ast}}p_{\ell}(\alpha^\ast)\}$ is a basis for $\mathbb{F}$ over $\mathbb{B}$, and assume $\{\mu_{1},\mu_{2},\dots,\mu_{\ell}\}$ is its dual basis. 
Then, $f(\alpha^{\ast})$ can be repaired by
\begin{equation}\label{dual basis equation}
f(\alpha^{\ast})=\sum\limits_{j=1}^{\ell}\mu_{j}tr_{\mathbb{F}/\mathbb{B}}(\upsilon_{\alpha^\ast}p_{j}(\alpha^\ast)f(\alpha^{\ast})).
\end{equation}
Since $\upsilon_{\alpha^{\ast}}$ is a non-zero constant, we equivalently suppose that $\{p_1(\alpha^\ast),\dots,p_\ell(\alpha^\ast)\}$ is a basis.

In fact, by \cite{guruswami2017repairing} any linear repair scheme of RS code for the failed node $f(\alpha^*)$ is equivalent to choosing $p_{j}(x), j \in [\ell],$ with degree smaller than $r$, such that $\{p_1(\alpha^\ast),\dots,p_\ell(\alpha^\ast)\}$ forms a basis for $\mathbb{F}$ over $\mathbb{B}$. We call this the {\bf full rank condition}:
\begin{align}\label{polynomial requirement}
rank_{\mathbb{B}}(\{p_{1}(\alpha^\ast),p_{2}(\alpha^\ast),\dots,p_{\ell}(\alpha^\ast)\})=\ell.
\end{align}

The repair bandwidth can be calculated from \eqref{eq:helper trans} and by noting that $v_i f(\alpha_i)$ is a constant:
\begin{align}\label{bandwidth calculation}
b=\sum\limits_{\alpha\in A,\alpha\neq \alpha^\ast} rank_{\mathbb{B}}(\{p_{1}(\alpha),p_{2}(\alpha),\dots,p_{\ell}(\alpha)\}).
\end{align}
We call this the {\bf repair bandwidth condition}.

The goal of a good RS code construction and its repair scheme is to choose appropriate evaluation points $A$ and polynomials $p_{j}(x), j \in [\ell],$ that can reduce the repair bandwidth in \eqref{bandwidth calculation} while satisfying \eqref{polynomial requirement}.

The following lemma is due to the structure of the multiplicative group of $\mathbb{F}$, which will be used for finding the evaluation points in the code constructions in this paper. Similar statements can be found in \cite[Ch. 2.6]{ryan2009channel}.

\begin{lem}\label{cyclic subgroups}
Assume $\mathbb{E} \le \mathbb{F}=GF(q^{\ell})$, then $\mathbb{F}^\ast$ can be partitioned to $t\triangleq \frac{q^\ell-1}{|\mathbb{E}|-1}$ cosets: $\{\mathbb{E}^\ast,\beta \mathbb{E}^\ast$, $\beta^2 \mathbb{E}^\ast,\dots,\beta^{t-1} \mathbb{E}^\ast\}$, where $\beta$ is a primitive element of $\mathbb{F}$.
\end{lem}

\begin{IEEEproof}
The \bl{$q^{\ell}-1$} elements in $\mathbb{F}^\ast$ are $\{1,\beta,\beta^2,\dots,\beta^{q^{\ell}-2}\}$  and $\mathbb{E}^\ast\subseteq \mathbb{F}^\ast$. Assume that $t$ is the smallest nonzero number that satisfies $\beta^t\in \mathbb{E}^\ast$, then we know that  $\beta^k\in \mathbb{E}^\ast$ if and only if $t|k$. Also,  $\beta^{k_{1}}\neq \beta^{k_{2}}$ when $k_{1}\neq k_{2}$ and $k_{1},k_{2}<q^\ell-2$. Since there are only $|\mathbb{E}|-1$ nonzero distinct elements in $\mathbb{E^\ast}$ and $\beta^{q^{\ell}-1}=1$, we have $t=\frac{q^\ell-1}{|\mathbb{E}|-1}$ and the $t$ cosets are $\mathbb{E}^\ast=\{1,\beta^t,\beta^{2t},\dots,\beta^{(|\mathbb{E}|-2)t}\}$,  
$\beta \mathbb{E}^\ast=\{\beta,\beta^{t+1},\beta^{2t+1},\dots,\beta^{(|\mathbb{E}|-2)t+1}\}$, $\dots,$ 
$\beta^{t-1} \mathbb{E}^\ast=$ $\{\beta^{t-1},$ $\beta^{2t-1},\beta^{3t+1},\dots,\beta^{(|\mathbb{E}|-1)t-1}\}$.
\end{IEEEproof}

\section{Reed-Solomon repair schemes for single erasure} 
\label{sec: single erasure}
In this section, we present our schemes in which  the evaluation points are part of one coset, two cosets and multiple cosets for a single erasure. From these constructions, we achieve several different points on the tradeoff between the sub-packetization size and the normalized repair bandwidth. The main ideas of the constructions are:

(i) In all our schemes, we take an original RS code, and construct a new code over a larger finite field. Thus, the sub-packetization size $\ell$ is increased.

(ii) For the schemes using one and two cosets, the code parameters $n, k$ are kept the same as the original code. Hence, for given $n,r=n-k$, the sub-packetization size $\ell$ increases, but we show that the normalized repair bandwidth remains the same.

(iii) For the scheme using multiple cosets, the code length $n$ is increased and the redundancy $r$ is fixed. Moreover, the code length $n$ grows faster than the sub-packetization size $\ell$. Therefore, for fixed $n,r$, the sub-packetization $\ell$ decreases, and we show that the normalized repair bandwidth is only slightly larger than the original code.

\subsection{Schemes in one coset}
Assume $\mathbb{E}=GF(q^a)$ is a subfield of $\mathbb{F}=GF(q^\ell)$ and $\mathbb{B}=GF(q)$ is the base field, where $q$ is a prime number. The evaluation points of the code over $\mathbb{F}$ that we construct are part of one coset in Lemma \ref{cyclic subgroups}. 

We first present the following lemma about the basis.
\begin{lem}\label{basis lemma1}
Assume $\{\xi_{1},\xi_{2},\dots,\xi_{\ell}\}$ is a basis for $\mathbb{F}=GF(q^\ell)$ over $\mathbb{B}=GF(q)$, then $\{\xi_{1}^{q^s},\xi_{2}^{q^s},\dots,\xi_{\ell}^{q^s}\}$, $s\in[\ell]$ is also a basis.
\end{lem}
\begin{IEEEproof}
Assume $\{\xi_{1}^{q^s},\xi_{2}^{q^s},\dots,\xi_{\ell}^{q^s}\}$, $s\in[\ell]$ is not a basis for $\mathbb{F}$ over $\mathbb{B}$, then there exist nonzero $(\alpha_1,\alpha_2,\dots,\alpha_\ell), \alpha_i \in\mathbb{B}, i \in [\ell],$ that satisfy
\begin{align}
&\alpha_1\xi_1^{q^s}+\alpha_2\xi_2^{q^s}+\dots+\alpha_\ell\xi_\ell^{q^s}\nonumber\\
=&0\nonumber\\
=&(\alpha_1\xi_1+\alpha_2\xi_2+\dots+\alpha_\ell\xi_\ell)^{q^s},
\end{align}
which is in contradiction to the assumption that $\{\xi_{1},\xi_{2},\dots,\xi_{\ell}\}$ is a basis for $\mathbb{F}$ over $\mathbb{B}$.
\end{IEEEproof}

The following theorem shows the repair scheme using one coset for the evaluation points.
\begin{thm}\label{scheme in one coset}
There exists an $RS(n,k)$ code over $\mathbb{F}=GF(q^\ell)$ with repair bandwidth $b \le \frac{\ell}{a}(n-1)(a-s)$ symbols over $\mathbb{B}=GF(q)$, where $q$ is a prime number and $a,s$ satisfy $n<q^a, q^s\leq n-k$, $a|\ell$. 
\end{thm}

\begin{IEEEproof}
Assume a field $\mathbb{F}=GF(q^\ell)$ is extended from $\mathbb{E}=GF(q^a)$, $a \mid \ell$, and  $\beta$ is a primitive element of $\mathbb{F}$. We focus on the code $RS(A,k)$ of dimension $k$ over $\mathbb{F}$ with evaluation points $A=\{\alpha_{1},\alpha_{2},\dots,\alpha_{n}\} \subseteq \beta^m\mathbb{E}^\ast$ for some $0 \leq m < \frac{q^\ell-1}{q^a-1}$, which is one of the cosets in Lemma \ref{cyclic subgroups}. The base field is $\mathbb{B}=GF(q)$ and (\ref{trace equation}) is used to repair the failed node $f(\alpha^\ast)$. 

\noindent\textbf{Construction I}:
Inspired by \cite{guruswami2017repairing}, for $s=a-1$, we choose
\begin{equation}\label{scheme1 polynomial for a-1}
p_{j}(x)=\frac{tr_{\mathbb{E}/\mathbb{B}}(\xi_{j}(\frac{x}{\beta^m}-\frac{\alpha^\ast}{\beta^m}))}{\frac{x}{\beta^m}-\frac{\alpha^\ast}{\beta^m}}, j\in[a],
\end{equation}
where $\{\xi_{1},\xi_{2},\dots,\xi_{a}\}$ is a basis for $\mathbb{E}$ over $\mathbb{B}$. The degree of $p_{j}(x)$ is smaller than $r$ since $q^s\leq r$. When $x=\alpha^\ast$, by \eqref{trace definition} we have
\begin{align}\label{eq:12}
p_{j}(\alpha^\ast)=\xi_{j}.
\end{align}
So, the polynomials satisfy
\begin{align}
rank_{\mathbb{B}}(\{p_{1}(\alpha^\ast),p_{2}(\alpha^\ast),\dots,p_{a}(\alpha^\ast)\})=a.
\end{align}
When $x\neq \alpha^\ast$, since $tr_{\mathbb{E}/\mathbb{B}}(\xi_{j}(\frac{x}{\beta^m}-\frac{\alpha^\ast}{\beta^m})) \in \mathbb{B}$, and $\frac{x}{\beta^m}-\frac{\alpha^\ast}{\beta^m}$ is a constant independent of $j$, we have
\begin{align}
rank_{\mathbb{B}}(\{p_{1}(x),p_{2}(x),\dots,p_{a}(x) \})= 1.
\end{align}
Let \{$\eta_1, \eta_2, \eta_3,\dots,\eta_{\ell/a}\}$ be a basis for $\mathbb{F}$ over $\mathbb{E}$, the $\ell$ repair polynomials are chosen as 
\begin{align}
\{\eta_1 p_{j}(x), \eta_2 p_{j}(x),\dots, \eta_{\ell/a}p_{j}(x):j\in [a]\}. 
\end{align}
Since $p_j(x)\in \mathbb{E}$,  we can conclude that
\begin{eqnarray}
rank_{\mathbb{B}}(\{\eta_1 p_{j}(\alpha^\ast), \eta_2 p_{j}(\alpha^\ast),\dots, \eta_{\ell/a}p_{j}(\alpha^\ast): j\in [a] \})\nonumber\\
=\frac{\ell}{a}rank_{\mathbb{B}}(\{p_{1}(\alpha^\ast),p_{2}(\alpha^\ast),\dots,p_{a}(\alpha^\ast) \})=\ell
\end{eqnarray}
satisfies the full rank condition, and for $x\neq \alpha^\ast$
\begin{align}
rank_{\mathbb{B}}(\{\eta_1 p_{j}(x), \eta_2 p_{j}(x),\dots, \eta_{\ell/a}p_{j}(x):j\in[a] \})\nonumber\\
=\frac{\ell}{a}rank_{\mathbb{B}}(\{p_{1}(x),p_{2}(x),\dots,p_{a}(x) \})=\frac{\ell}{a}.
\end{align}
From \eqref{bandwidth calculation} we can calculate the repair bandwidth
\begin{equation}\label{repair bandwidth of construction 1}
b=\frac{\ell}{a}(n-1).
\end{equation}

\noindent\textbf{Construction II}: For $s\leq a-1$, inspired by \cite{dau2017optimal}, we choose 
\begin{equation}\label{scheme1 polynomial}
p_{j}(x)=\xi_{j}\prod\limits_{i=1}^{q^s-1}\left(\frac{x}{\beta^m}-\left(\frac{\alpha^\ast}{\beta^m}-w_i^{-1}\xi_{j}\right)\right), j\in[a],
\end{equation}
where $\{\xi_{1},\xi_{2},\dots,\xi_{a}\}$ is a basis for $\mathbb{E}$ over $\mathbb{B}$, and $W=\{w_0=0,w_{1},w_{2},\dots,w_{q^s-1}\}$ is an $s$-dimensional subspace in $\mathbb{E}$, $s<a,q^s\leq r$. It is easy to check that the degree of $p_{j}(x)$ is smaller than $r$ since $q^s \le r$. When $x=\alpha^\ast$, we have
\begin{align} \label{eq:20}
p_{j}(\alpha^\ast)=\xi_{j}^{q^s}\prod\limits_{i=1}^{q^s-1}w_i^{-1}.
\end{align}
Since $\prod\limits_{i=1}^{q^s-1}w_i^{-1}$ is a constant, from Lemma \ref{basis lemma1} we have
\begin{align}
rank_{\mathbb{B}}(\{p_{1}(\alpha^\ast),p_{2}(\alpha^\ast),\dots,p_{a}(\alpha^\ast) \})=a.
\end{align}

For $x\neq \alpha^\ast$, set $x'=\frac{\alpha^\ast}{\beta^m}-\frac{x}{\beta^m} \in \mathbb{E}$, we have
\begin{align}
p_{j}(x)&=\xi_{j}\prod\limits_{i=1}^{q^s-1}\left(\frac{x}{\beta^m}-\left(\frac{\alpha^\ast}{\beta^m}-w_i^{-1}\xi_{j}\right)\right)\nonumber\\
&=\xi_{j}\prod\limits_{i=1}^{q^s-1}(w_i^{-1}\xi_{j}-x')\nonumber\\
&=\xi_{j}\prod\limits_{i=1}^{q^s-1}(w_i^{-1}x')\prod\limits_{i=1}^{q^s-1}(\xi_{j}/x'-w_i)\nonumber\\
&=(x')^{q^s}\prod\limits_{i=1}^{q^s-1}(w_i^{-1})\prod\limits_{i=0}^{q^s-1}(\xi_{j}/x'-w_i). \label{eq:22}
\end{align}
By \cite[p. 4]{goss2012basic}, $g(y)=\prod\limits_{i=0}^{q^s-1}(y-w_i)$ is a linear mapping from $\mathbb{E}$ to itself with dimension $a-s$ over $\mathbb{B}$. Since $(x')^{q^s}\prod\limits_{i=1}^{q^s-1}(w_i^{-1})$ is a constant independent of $j$, we have
\begin{align}
rank_{\mathbb{B}}(\{p_{1}(x),p_{2}(x),\dots,p_{a}(x) \})\leq a-s.
\end{align}
Let \{$\eta_1, \eta_2, \eta_3,\dots,\eta_{\ell/a}\}$ be a basis for $\mathbb{F}$ over $\mathbb{E}$, then the $\ell$ polynomials are chosen as $\{\eta_1 p_{j}(x)$, $\eta_2 p_{j}(x),\dots, \eta_{\ell/a}p_{j}(x),j\in [a]\}$. From \eqref{eq:20} and \eqref{eq:22} we know that $p_j(x)\in \mathbb{E}$, so we can conclude that
\begin{eqnarray}\label{rank3}
rank_{\mathbb{B}}(\{\eta_1 p_{j}(\alpha^\ast), \eta_2 p_{j}(\alpha^\ast),\dots, \eta_{\ell/a}p_{j}(\alpha^\ast): j\in [a] \})\nonumber \\
=\frac{\ell}{a}rank_{\mathbb{B}}(\{p_{1}(\alpha^\ast),p_{2}(\alpha^\ast),\dots,p_{a}(\alpha^\ast) \})=\ell
\end{eqnarray}
satisfies (\ref{polynomial requirement}), and for $x\neq \alpha^\ast$
\begin{align}\label{rank4}
rank_{\mathbb{B}}(\{\eta_1 p_{j}(x), \eta_2 p_{j}(x),\dots, \eta_{\ell/a}p_{j}(x): j\in[a] \})\nonumber \\
=\frac{\ell}{a}rank_{\mathbb{B}}(\{p_{1}(x),p_{2}(x),\dots,p_{a}(x) \})\le\frac{\ell}{a}(a-s).
\end{align}
Now from (\ref{bandwidth calculation}) we can calculate the repair bandwidth 
\begin{equation}\label{repair bandwidth of construction 2}
 b\le\frac{\ell}{a}(n-1)(a-s).
\end{equation}
Combining \eqref{repair bandwidth of construction 1} and \eqref{repair bandwidth of construction 2} will complete the proof of Theorem \ref{scheme in one coset}.
\end{IEEEproof}

Rather than directly using the schemes in \cite{guruswami2017repairing} and \cite{dau2017optimal}, the polynomials \eqref{scheme1 polynomial for a-1} and \eqref{scheme1 polynomial} that we use are similar to \cite{guruswami2017repairing} and \cite{dau2017optimal}, respectively, but are mappings from $\mathbb{E}$ to $\mathbb{B}$. Moreover, we multiply each polynomial with the basis for $\mathbb{F}$ over $\mathbb{E}$ to  satisfy the full rank condition. In this case, our scheme significantly reduces the repair bandwidth when the code length remains the same. Our evaluation points are in a coset rather than the entire field $\mathbb{F}$ as in \cite{guruswami2017repairing} and \cite{dau2017optimal}. It should be noted that $a$ here can be an arbitrary number that divides $\ell$ and when $a=\ell$, our schemes are exactly the same as those in \cite{guruswami2017repairing} and \cite{dau2017optimal}. Note that the normalized repair bandwidth $\frac{b}{\ell (n-1)}$ decreases as $a$ decreases. Therefore, our scheme outperforms those in \cite{guruswami2017repairing} and \cite{dau2017optimal} when applied to the case of $\ell > \log_q n$.

\begin{exa}
Assume $q=2, \ell=9, a=3$ and $\mathbb{E}=\{0, 1, \alpha, \alpha^2,\dots, \alpha^6\}$. Let $A=\mathbb{E}^\ast$, $n=7,k=5$ so $r=n-k=2$. Choose $s=\log_2r=1$ and $W=\{0, 1\}$ in Construction II.  Then, we have $p_j(x)=\xi_j(x-\alpha^\ast+\xi_j)$. Let $\{\xi_{1},\xi_{2},\xi_{3}\}$ be $\{1, \alpha, \alpha^2\}$. It is easy to check that $rank_{\mathbb{B}}(\{p_{1}(\alpha^\ast),p_{2}(\alpha^\ast),p_{3}(\alpha^\ast)\})=3$ and $rank_{\mathbb{B}}(\{p_{1}(x),p_{2}(x),p_{3}(x)\})=2$ for $x\neq \alpha^\ast$. Therefore the repair bandwidth is $b=36$ bits as suggested in Theorem \ref{scheme in one coset}. For the same $(n,k,\ell)$, the repair bandwidth in \cite{dau2017optimal} is $48$ bits. For another example, consider $RS(14,10)$ code used in Facebook \cite{sathiamoorthy2013xoring}, we have repair bandwidth of $52$ bits for $\ell=8$, while \cite{dau2017optimal} requires $60$ bits and the naive scheme requires $80$ bits.
\end{exa}

\begin{rem}
The schemes in \cite{guruswami2017repairing} and \cite{dau2017optimal} can also be used in an RS code over $\mathbb{E}$ with repair bandwidth $(n-1)(a-s)$, and with $\ell/a$ copies of the code. Thus, they can also reach the repair bandwidth of $\frac{\ell}{a}(n-1)(a-s)$. It should be noted that by doing so, the code is a vector code, however our scheme constructs a scalar code. To the best of our knowledge, this is the first example of such a scalar code in the literature.
\end{rem}

\subsection{Schemes in two cosets}
Now we discuss our scheme when the evaluation points are chosen from two cosets. In this scheme, we choose the polynomials that have full rank when evaluated at the coset containing the failed node, and rank $1$ when evaluated at the other coset. 

\begin{thm}
There exists an $RS(n,k)$ code over $\mathbb{F}=GF(q^\ell)$ with repair bandwidth $b<(n-1)\frac{\ell+a}{2}$ symbols over $\mathbb{B}=GF(q)$, where $q$ is a prime number and $a$ satisfies $n\leq2(q^a-1)$, $a | \ell$, $\frac{\ell}{a}\leq n-k$.
\end{thm}

\begin{IEEEproof}
Assume a field $\mathbb{F}=GF(q^\ell)$ is extended from $\mathbb{E}=GF(q^a)$ and $\beta$ is the primitive element of $\mathbb{F}$. We focus on the code $RS(A,k)$ over $\mathbb{F}$ of dimension $k$ with evaluation points $A$ consisting of $n/2$ points from $\beta^{m_1}\mathbb{E}^\ast$ and $n/2$ points from $\beta^{m_2}\mathbb{E}^\ast$, $0\leq m_1<m_2\leq\frac{q^\ell-1}{q^a-1}$ and $m_2-m_1=q^s$, $s\in \{0,1,\dots,\frac{\ell}{a}\}$. 

In this case we view $\mathbb{E}$ as the base field and repair the failed node $f(\alpha^\ast)$ by
\begin{equation}\label{trace equation2}
tr_{\mathbb{F}/\mathbb{E}}(\upsilon_{\alpha^{\ast}}p_{j}(\alpha^{\ast})f(\alpha^{\ast}))=-\sum\limits_{i=1,\alpha_i\neq\alpha^{\ast}}^{n}tr_{\mathbb{F}/\mathbb{E}}(\upsilon_{i}p_{j}(\alpha_{i})f(\alpha_{i})).
\end{equation}

Inspired by \cite[Theorem 10]{guruswami2017repairing}, for $j\in[\frac{\ell}{a}]$, we choose
\begin{equation}
p_{j}(x)=
\begin{cases}
(\frac{x}{\beta^{m_{2}}})^{j-1},\text{ if }  \alpha^\ast \in \beta^{m_{1}}\mathbb{E}^\ast,\\
(\frac{x}{\beta^{m_{1}}})^{j-1},\text{ if }  \alpha^\ast \in \beta^{m_{2}}\mathbb{E}^\ast.
\end{cases}
\end{equation}
The degree of $p_{j}(x)$ is smaller than $r$ when $\frac{\ell}{a}\leq r$. Then, we check the rank in each case.

When $\alpha^\ast \in \beta^{m_{2}}\mathbb{E}^\ast$, if $x = \beta^{m_{1}}\gamma \in\beta^{m_{1}}\mathbb{E}^\ast$, for some $\gamma\in\mathbb{E}^\ast$,
\begin{align}
p_{j}(x)&=\left(\frac{x}{\beta^{m_{1}}}\right)^{j-1}=\gamma^{j-1},
\end{align}
 so
\begin{align}
rank_{\mathbb{E}}(\{p_{1}(x),p_{2}(x),\dots,p_{\frac{\ell}{a}}(x) \})=1.
\end{align}
If $x = \beta^{m_{2}}\gamma \in\beta^{m_{2}}\mathbb{E}^\ast$, for some $\gamma\in\mathbb{E}^\ast$,
\begin{align}
p_{j}(x)&=\left(\frac{x}{\beta^{m_{1}}}\right)^{j-1}=(\beta^{m_2-m_1})^{j-1}\gamma^{j-1}.
\end{align}
Since $m_2-m_1=q^s$ and $\{1,\beta,\beta^2,\dots,\beta^{\frac{\ell}{a}-1}\}$ is the polynomial basis for $\mathbb{F}$ over $\mathbb{E}$, from Lemma \ref{basis lemma1} we know that
\begin{align}
rank_{\mathbb{E}}(\{p_{1}(x),p_{2}(x),\dots,p_{\frac{\ell}{a}}(x) \})=\frac{\ell}{a}.
\end{align}

When $\alpha^\ast \in \beta^{m_{1}}\mathbb{E}^\ast$, if $x = \beta^{m_{1}}\gamma \in\beta^{m_{1}}\mathbb{E}^\ast$, for some $\gamma\in\mathbb{E}^\ast$,
\begin{align}
p_{j}(x)&=\left(\frac{x}{\beta^{m_{2}}}\right)^{j-1}\nonumber\\
&=(\beta^{m_1-m_2})^{j-1}\gamma^{j-1}\nonumber\\
&=(\beta^{m_2-m_1})^{1-\frac{\ell}{a}}(\beta^{m_2-m_1})^{\frac{\ell}{a}-j}\gamma^{j-1}.
\end{align}
Since $(\beta^{m_2-m_1})^{1-\frac{\ell}{a}}$ is a constant, from Lemma \ref{basis lemma1} we know that
\begin{align}
rank_{\mathbb{E}}(\{p_{1}(x),p_{2}(x),\dots,p_{\frac{\ell}{a}}(x) \})=\frac{\ell}{a}.
\end{align}
If $x = \beta^{m_{2}}\gamma \in\beta^{m_{2}}\mathbb{E}^\ast$, for some $\gamma\in\mathbb{E}^\ast$,
\begin{align}
p_{j}(x)&=\left(\frac{x}{\beta^{m_{2}}}\right)^{j-1}=\gamma^{j-1},
\end{align}
so
\begin{align}
rank_{\mathbb{E}}(\{p_{1}(x),p_{2}(x),\dots,p_{\frac{\ell}{a}}(x) \})=1.
\end{align}

Therefore, $\{p_j(\alpha^\ast),j\in [\frac{\ell}{a}]\}$ has full rank over $\mathbb{E}$, for any evaluation point $\alpha^\ast \in A$. For $x$ from the coset containing $\alpha^\ast$, the polynomials have rank $\ell/a$, and for $x$ from the other coset, the polynomials have rank $1$. Then, the repair bandwidth in symbols over $\mathbb{B}$ can be calculated from (\ref{bandwidth calculation}) as
\begin{align} \label{eq: bandwidth two cosets}
b&=\frac{\ell}{a}(\frac{n}{2}-1)\log_q|\mathbb{E}|+\frac{n}{2}\log_q|\mathbb{E}|\nonumber \\
&=(n-1)\frac{\ell+a}{2}-\frac{\ell-a}{2}\nonumber \\
&<(n-1)\frac{\ell+a}{2}.
\end{align}
Thus, the proof is completed.
\end{IEEEproof}

\begin{exa}
Take the $RS(14,11)$ code over $\mathbb{F}=GF(2^{12})$ for example. Let $\beta$ be the primitive element in $\mathbb{F}$, $a=4$, $s=\ell/a=3$ and $A=\mathbb{E}^\ast \cup \beta\mathbb{E}^\ast$. Assume $\alpha^\ast \in \beta\mathbb{E}^\ast$, then $\{p_j(x), j\in[3] \}$ is the set $\{1, x, x^2 \}$. It is easy to check that when  $x \in \beta\mathbb{E}^\ast$ the polynomials have full rank and  when  $x \in \mathbb{E}^\ast$ the polynomials have rank $1$. The total repair bandwidth is $100$ bits. For the same $(n,k,\ell)$, the repair bandwidth of our scheme in one coset is $117$ bits. For the scheme in \cite{guruswami2017repairing}, which only works for $\ell/a=2$, we can only choose $a=6$ and get the repair bandwidth of $114$ bits for the same $(n,k,\ell)$.
\end{exa}

\subsection{Schemes in multiple cosets}
\label{sec: 1 erasure multi cosets}
In the schemes in this subsection, we extend an original code to a new code over a larger field and the evaluation points are chosen from multiple cosets in Lemma \ref{cyclic subgroups} to increase the code length. The construction ensures that for fixed $n$, the sub-packetization size is smaller than the original code.
If the original code satisfies  several conditions to be discussed soon, the repair bandwidth in the new code is only slightly larger than that of the original code. Particularly, if the original code is an MSR code, then we can get the new code in a much smaller sub-packetization level with a small extra repair bandwidth. Also, if the original code works for any number of helpers and multiple erasures, the new code works for any number of helpers and multiple erasures, too. We discuss multiple erasures in Section \ref{sec:multiple erasures}.


We first prove a lemma regarding the ranks over different base fields, and then describe the new code.

\begin{lem}\label{ranks for extended fields}
Let $\mathbb{B}=GF(q), \mathbb{F}'=GF(q^{\ell'}), \mathbb{E}=GF(q^a)$, $\mathbb{F}=GF(q^{\ell}), \ell=a \ell'$. $a$ and $\ell'$ are relatively prime and $q$ can be any power of a prime number. For any set of $\{\gamma_1,\gamma_2,...,\gamma_{\ell'}\} \subseteq \mathbb{F'} \le \mathbb{F}$, we have  
\begin{align}
&rank_{\mathbb{E}}(\{\gamma_1,\gamma_2,...,\gamma_{\ell'} \})\nonumber\\
=&rank_{\mathbb{B}}(\{\gamma_1,\gamma_2,...,\gamma_{\ell'}\}).
\end{align}
\end{lem}
\begin{IEEEproof}
Assume $rank_{\mathbb{B}}(\{\gamma_1,\gamma_2,...,\gamma_{\ell'}\})=c$ and without loss of generality, $\{\gamma_1,\gamma_2,...,\gamma_c\}$ are linearly independent over $\mathbb{B}$. Then, we can construct $\{\gamma'_{c+1},\gamma'_{c+2},...,\gamma'_{\ell'}\}\subseteq\mathbb{F'}$  to make $\{\gamma_1,\gamma_2,...,\gamma_c,\gamma'_{c+1},$  $\gamma'_{c+2},...,\gamma'_{\ell'}\}$ form a basis for $\mathbb{F'}$ over $\mathbb{B}$.

Assume we get $\mathbb{F}$ by adjoining $\beta$ to $\mathbb{B}$. Then, from \cite[Theorem 1.86]{lidl1994introduction} we know that $\{1,\beta,\beta^2,...,\beta^{\ell'-1}\}$ is a basis for both $\mathbb{F}$ over $\mathbb{E}$, and $\mathbb{F'}$ over $\mathbb{B}$. So, any symbol $y\in\mathbb{F}$ can be presented as a linear combination of $\{1,\beta,\beta^2,...,\beta^{\ell'-1}\}$ with some coefficients in $\mathbb{E}$. Also, we know that there is an invertible linear transformation with coefficients in $\mathbb{B}$ between $\{\gamma_1,\gamma_2,...,\gamma_c,\gamma'_{c+1},\gamma'_{c+2},...,\gamma'_{\ell'}\}$ and $\{1,\beta,\beta^2,...,\beta^{\ell'-1}\}$, because they are a basis for $\mathbb{F}'$ over $\mathbb{B}$. Combined with the fact that $\{1,\beta,\beta^2,...,\beta^{\ell'-1}\}$ is also a basis for $\mathbb{F}$ over $\mathbb{E}$, we can conclude that any symbol $y\in\mathbb{F}$ can be represented as
\begin{align}
y=x_1\gamma_1+x_2\gamma_2+...+x_c\gamma_c+x_{c+1}\gamma'_{c+1}+...+x_{\ell}\gamma'_{\ell'}
\end{align}
with some coefficients $x_i\in\mathbb{E}$, which means 
that $\{\gamma_1,\gamma_2,...,\gamma_c,\gamma'_{c+1},\gamma'_{c+2},...,\gamma'_{\ell'}\}$ is also a basis for $\mathbb{F}$ over $\mathbb{E}$. Then, we have that $\{\gamma_1,\gamma_2,...,\gamma_c\}$ are linearly independent over $\mathbb{E}$, 
\begin{align}
&rank_{\mathbb{E}}(\{\gamma_1,\gamma_2,...,\gamma_{\ell'} \})\nonumber\\
\geq&c\nonumber\\
=&rank_{\mathbb{B}}(\{\gamma_1,\gamma_2,...,\gamma_{\ell'}\}).
\end{align}
Since $\mathbb{B} \le \mathbb{E}$, we also have
\begin{align}
&rank_{\mathbb{E}}(\{\gamma_1,\gamma_2,...,\gamma_{\ell'}\})\nonumber\\
\leq&rank_{\mathbb{B}}(\{\gamma_1,\gamma_2,...,\gamma_{\ell'}\}).
\end{align}
The proof is completed. 
\end{IEEEproof}

\begin{thm}\label{thm:bandwidth multiple cosets}
Assume there exists a $RS(n',k')$ code $\mathscr{E}'$ over $\mathbb{F'}=GF(q^{\ell'})$ with evaluation points set $A'$. The evaluation points are linearly independent over $\mathbb{B}=GF(q)$. The repair bandwidth is $b'$ and the repair polynomials are $p'_j(x)$. Then, we can construct a new $RS(n,k)$ code $\mathscr{E}$ over $\mathbb{F}=GF(q^{\ell}),\ell=a\ell'$ with $n= (q^a-1)n',k=n-n'+k'$ and repair bandwidth of $b= ab'(q^a-1)+(q^a-2)\ell$ symbols over $\mathbb{B}=GF(q)$ if we can find new repair polynomials $p_j(x)\in\mathbb{F}[x], j \in [\ell'],$ with degrees less than $n-k$ that satisfy
\begin{align}\label{new repair condition}
&rank_{\mathbb{E}}(\{p_{1}(x),p_{2}(x),\dots,p_{\ell'}(x) \})\nonumber\\
=&rank_{\mathbb{B}}(\{p'_{1}(\alpha),p'_{2}(\alpha),\dots,p'_{\ell'}(\alpha)\})
\end{align}
for all $\alpha \in A', x\in \alpha \mathbb{E}^\ast$, where $\mathbb{E}=GF(q^a)$.
\end{thm}

\begin{IEEEproof}
We first prove the case when $a$ and $\ell'$ are necessarily relatively prime using Lemma \ref{ranks for extended fields}, the case when $a$ and $\ell'$ are not relatively prime are proved in Appendix A.
Assume the evaluation points of $\mathscr{E}'$ are $A'=\{\alpha_1,\alpha_2,\dots,\alpha_{n'} \}$, then from Lemma \ref{ranks for extended fields} we know that they are also linearly independent over $\mathbb{E}$, so there does not exist $\gamma_i,\gamma_j\in\mathbb{E}^\ast$ that satisfy $\alpha_i\gamma_i=\alpha_j\gamma_j$, which implies that $\{\alpha_1\mathbb{E}^\ast, \alpha_2\mathbb{E}^\ast,\dots,\alpha_{n'}\mathbb{E}^\ast\}$ are distinct cosets. Then, we can extend the evaluation points to be 
\begin{align}\label{extended evaluation points}
A=\{\alpha_1\mathbb{E}^\ast, \alpha_2\mathbb{E}^\ast,\dots,\alpha_{n'}\mathbb{E}^\ast\}. 
\end{align}
and $n=(q^a-1)n'$. We keep the same redundancy $r=n'-k'$ for the new code so $k=n-r$.

For the new code $\mathscr{E}$, we use $p_j(x)\in\mathbb{F}[x], j \in [\ell']$ to repair the failed node $f(\alpha^\ast)$
\begin{align}\label{trace equation4}
&tr_{\mathbb{F}/\mathbb{E}}(\upsilon_{\alpha^{\ast}}p_{j}(\alpha^{\ast})f(\alpha^{\ast}))=-\sum\limits_{\alpha \in A, \alpha \neq\alpha^{\ast}} tr_{\mathbb{F}/\mathbb{E}}(\upsilon_{\alpha}p_{j}(\alpha)f(\alpha)).
\end{align}

Assume the failed node is $f(\alpha^\ast)$ and $\alpha^\ast\in\alpha_i \mathbb{E}^\ast$. Then, for the node $x\in\alpha_i \mathbb{E}^\ast$, because the original code satisfies the full rank condition, we have
\begin{align}
&rank_{\mathbb{E}}(\{p_{1}(x),p_{2}(x),\dots,p_{\ell'}(x) \})\nonumber\\
=&rank_{\mathbb{B}}(\{p'_{1}(\alpha_i),p'_{2}(\alpha_i),\dots,p'_{\ell'}(\alpha_i)\})=\ell',
\end{align}
then we can recover the failed node with $p_j(x)$, and each helper in the coset containing the failed node transmits $\ell'$ symbols over $\mathbb{E}$.

For a helper in the other cosets, $x\in\alpha_\epsilon \mathbb{E}^\ast, \epsilon\neq i$, by \eqref{new repair condition},
\begin{align}
&rank_{\mathbb{E}}(\{p_{1}(x),p_{2}(x),\dots,p_{\ell'}(x) \})\nonumber\\
=&rank_{\mathbb{B}}(\{p'_{1}(\alpha_\epsilon),p'_{2}(\alpha_\epsilon),\dots,p'_{\ell'}(\alpha_\epsilon)\}),
\end{align}
then every helper in these cosets transmits $\frac{b'}{n'-1}$ symbols in $\mathbb{E}$ on average. 

The repair bandwidth of the new code can be calculated from the repair bandwidth condition \eqref{bandwidth calculation} as
\begin{align}\label{new repair bandwidth}
b&=\frac{b'}{n'-1}\cdot(n'-1)|\mathbb{E}^\ast|\cdot a+(|\mathbb{E}^\ast|-1)\ell' \cdot a\nonumber\\
&=ab'(q^a-1)+(q^a-2)\ell
\end{align}
which completes the proof.
\end{IEEEproof}

Note that the calculation in \eqref{new repair bandwidth} and \eqref{eq: bandwidth two cosets} are similar in the sense that a helper in the coset containing the failure naively transmits the entire stored information, and the other helpers use the bandwidth that is the same as the original code.

As a special case of Theorem \ref{thm:bandwidth multiple cosets}, when $b'=\frac{\ell'}{r}(n'-1)$ matching the MSR bound \eqref{MSR bound}, we get
\begin{equation}
b=\frac{\ell}{r}(n-1)+\frac{\ell}{r}(r-1)(q^a-2),
\end{equation}
where the second term is the extra bandwidth compared to the MSR bound.

Next, we apply  Theorem \ref{thm:bandwidth multiple cosets} to the near-MSR code \cite{ye2016explicit} and the MSR code \cite{tamo2017optimal}. The first realization of the scheme in multiple cosets is inspired by \cite{ye2016explicit}.
\begin{thm}\label{thm:2nd multiple cosets}
There exists an $RS(n,k)$ code over $\mathbb{F}=GF(q^{\ell})$ of which $n=(q^a-1)\log_r\frac{\ell}{a}$ and $a|\ell$, such that the repair bandwidth satisfies $b<\frac{\ell}{n-k}[n+1+(n-k-1)(q^a-2)]$, measured in symbols over $\mathbb{B}=GF(q)$ for some prime number $q$. 
\end{thm}

\begin{IEEEproof}
We first prove the case when $a$ and $\ell'$ are relatively prime using Lemma \ref{ranks for extended fields}, the case when $a$ and $\ell'$ are not necessarily relatively prime are proved in Appendix A. We use the code in \cite{ye2016explicit} as the original code. The original code is defined in $\mathbb{F'}=GF(q^{\ell'})$ and $\ell'=r^{n'}$. The evaluation points are $A'=\{\beta ,\beta^{r} ,\beta^{r^{2}} ,\dots,\beta^{r^{n'-1}} \}$ where $\beta$ is a primitive element of $\mathbb{F'}$. 

In the original code, for $c=0,1,2,\dots,\ell'-1$, we write its $r$-ary expansion as $c=(c_{n'}c_{n'-1}\dots c_{1})$, where $0\leq c_{i} \leq r-1$ is the $i$-th digit from the right. Assuming the failed node is $ f(\beta^{r^{i-1}})$, the repair polynomials are chosen to be
\begin{equation}
p'_{j}(x)=\beta^c x^s, c_{i}=0,s=0,1,2,\dots,r-1, x\in\mathbb{F'}.
\end{equation}
Here $c$ varies from $0$ to $\ell'-1$ given that $c_{i}=0$, and $s$ varies from $0$ to $r-1$. So, we have $\ell'$ polynomials in total. The subscript $j$ is indexed by $c$ and $s$, and by a small abuse of the notation, we write $j\in[\ell']$.

In the new code, let us define $\mathbb{E}=GF(q^a)$ of which $a$ and $\ell'$ are relatively prime. Adjoining $\beta$ to $\mathbb{E}$, we get $\mathbb{F}=GF(q^{\ell}),\ell=a\ell'$.
The new evaluation points are $A=\{\beta \mathbb{E}^\ast,\beta^{r} \mathbb{E}^\ast,\beta^{r^{2}} \mathbb{E}^\ast,\dots,\beta^{r^{n'-1}} \mathbb{E}^\ast\}$.
Since $A'$ is part of the polynomial basis for $\mathbb{F'}$ over $\mathbb{B}$, we know that $\{\beta ,\beta^{r} ,\beta^{r^{2}} ,\dots,\beta^{r^{n'-1}} \}$ are linearly independent over $\mathbb{B}$. Hence, we can apply Lemma \ref{ranks for extended fields} and the cosets are distinct, resulting in $n=|A|=(q^a-1)\log_r\frac{\ell}{a}$.

In our new code, let us assume the failed node is $f(\alpha^\ast)$ and $\alpha^\ast \in \beta^{r^{i-1}} C$, and we choose the polynomial $p_{j}(x)$ with the same form as $p'_{j}(x)$,
\begin{equation}
p_{j}(x)=\beta^c x^s, c_{i}=0,s=0,1,2,\dots,r-1,x\in\mathbb{F}.
\end{equation}

For nodes corresponding to $x = \beta^{r^t}\gamma \in \beta^{r^t} \mathbb{E}^\ast$, for some $\gamma \in \mathbb{E}^\ast$, we know that
\begin{equation}
p_{j}(x)=\beta^c x^s=\beta^c (\gamma\beta^{r^t})^s=\gamma^sp'_{j}(\beta^{r^t}).
\end{equation}
Since $p'_{j}(\beta^{r^t})\in \mathbb{F'}$, from Lemma \ref{ranks for extended fields}, we have
\begin{align}
&rank_{\mathbb{E}}(\{\gamma^sp'_{1}(\beta^{r^t}),\gamma^sp'_{2}(\beta^{r^t}),\dots,\gamma^sp'_{\ell'}(\beta^{r^t}) \})\nonumber\\
=&rank_{\mathbb{E}}(\{p'_{1}(\beta^{r^t}),p'_{2}(\beta^{r^t}),\dots,p'_{\ell'}(\beta^{r^t}) \})\nonumber\\
=&rank_{\mathbb{B}}(\{p'_{1}(\beta^{r^t}),p'_{2}(\beta^{r^t}),\dots,p'_{\ell'}(\beta^{r^t})\}),
\end{align}
which satisfies \eqref{new repair condition}. Since the repair bandwidth of the original code is $b'<(n'+1)\frac{\ell'}{r}$, from \eqref{new repair bandwidth} we can calculate the repair bandwidth as
\begin{align}
b&=ab'(q^a-1)+(q^a-2)\ell \nonumber\\
&<\frac{\ell}{r}[n+1+(r-1)(q^a-2)],
\end{align}
where the second term is the extra bandwidth compared to the original code.
\end{IEEEproof}

\begin{exa}
We take an $RS(4,2)$ code in $GF(2^{16})$ as the original code and extend it with $a=3, |\mathbb{E}^\ast|=7$ to an $RS(28,26)$ code in $GF(2^{48 })$ with normalized repair bandwidth of $\frac{b}{(n-1)\ell}<0.65$. The $RS(28,26)$ code in \cite{ye2016explicit} achieves the normalized repair bandwidth of $\frac{b}{(n-1)\ell}<0.54$, while it requires $\ell=2.7\times 10^8$. Our scheme has a much smaller $\ell$ compared to the scheme in \cite{ye2016explicit} while the repair bandwidth is a bit larger.  
\end{exa}

In the above theorem, we extend \cite{ye2016explicit} to a linearly larger sub-packetization and an exponentially larger code length, which means that for the same code length, we can have a much smaller sub-packetization level.


Next, we show our second realization of the scheme in multiple cosets, which is inspired by \cite{tamo2017optimal}. Different from the previous constructions, this one allows any number of helpers, $k \le d \le n-1$. The sub-packetization size in the original code of \cite{tamo2017optimal} satisfies $\ell' \approx ({n'})^{n'}$ when $n'$ grows to infinity, thus in our new code it satisfies $\ell \approx a ({n'})^{n'}$ for some integer $a$.


%
%

\begin{thm}\label{second realization of multiple cosets}
Let $q$ be a prime number. There exists an $RS(n,k)$ code over $\mathbb{F}=GF(q^{\ell})$ of which $\ell=asq_1q_2...q_{\frac{n}{q^a-1}}$, where $q_i$ is the  $i$-th prime number that satisfies $s|(q_i-1),s = d-k+1$ and $a$ is some integer. $d$ is the number of helpers, $k\leq d\leq (n-1)$. The average repair bandwidth is $b=\frac{d\ell}{(n-1)(d-k+1)}[n-1+(d-k)(q^a-2)]$ measured in symbols over $\mathbb{B}=GF(q)$.
\end{thm}

\begin{IEEEproof}
We first prove the case when $a$ and $\ell'$ are relatively prime using Lemma \ref{ranks for extended fields}, the case when $a$ and $\ell'$ are not necessarily relatively prime are proved in Appendix A. We use the code in \cite{tamo2017optimal} as the original code, where the number of helpers is $d'$. We set $n-k=n'-k'$ and calculate the repair bandwidth for $d$ helpers from the original code when $d'=d-k+k'$. Let us define $\mathbb{F}_{q}(\alpha)$ to be the field obtained by adjoining $\alpha$ to the base field $\mathbb{B}$. Similarly, we define  $\mathbb{F}_{q}(\alpha_1,\alpha_2,\dots,\alpha_n)$ for adjoining multiple elements. Let $\alpha_i$ be an element of order $q_i$ over $\mathbb{B}$. The code is defined in the field $\mathbb{F'}=GF(q^{\ell'})=GF(q^{sq_1q_2\dots,q_{n'}})$, which is the degree-s extension of $\mathbb{F}_{q}(\alpha_1, \alpha_2, \dots ,\alpha_{n'})$. The evaluation points are $A'=\{\alpha_1, \alpha_2, \dots, \alpha_{n'}\}$. 

Assuming the failed node is $f(\alpha_i)$ and the helpers are chosen from the set $R'$, $|R'|=d'$, the base field for repair is $\mathbb{F'}_i$, defined as $\mathbb{F'}_i\triangleq \mathbb{F}_{q}(\alpha_j, j\in[n'], j\neq i)$.
The repair polynomials are $\{\eta_t p'_j(\alpha_i), t \in [q_i], j \in [s]\}$, where
\begin{align}
p'_{j}(x)=x^{j-1}g'(x), j\in[s],x\in\mathbb{F'},
\end{align}
\begin{align}
g'(x)=\prod_{\alpha \in A/ (R'\cup\{\alpha_i\} ) } (x-\alpha), x\in\mathbb{F'}.
\end{align}
and $\eta_t\in \mathbb{F'}, t \in [q_i],$ are constructed in \cite{tamo2017optimal} such that $\{\eta_t p'_j(\alpha_i), t \in [q_i], j \in [s]\}$ forms the basis for $\mathbb{F'}$ over $\mathbb{F'}_i$. 
The repair is done using
\begin{align}
&tr_{\mathbb{F'}/\mathbb{F'}_i}(\upsilon_{\alpha_i}\eta_tp'_{j}(\alpha_i)f'(\alpha_i))=-\sum\limits_{\epsilon=1,\epsilon\neq i}^{n'}tr_{\mathbb{F'}/\mathbb{F'}_i}(\upsilon_{\epsilon}\eta_tp'_{j}(\alpha_{\epsilon})f'(\alpha_{\epsilon})).
\end{align}
For $x\notin R'\cup\{\alpha_i\}$, $p'_j(x)=0$, so no information is transmitted. The original code reaches the MSR repair bandwidth 
\begin{align}
b'&=\sum\limits_{\epsilon\in R'}rank_{\mathbb{F'}_{i}}(\{\eta_tp'_{j}(\alpha_{\epsilon}): t \in [q_i], j \in [s]\})\nonumber\\
&=\frac{d'\ell'}{d'-k'+1}.
\end{align}

In our new code, we define $\mathbb{E}=GF(q^a)=\mathbb{F}_{q}(\alpha_{n+1})$ where $a$ and $\ell'$ are relatively prime, and $\alpha_{n+1}$ is an element of order $a$ over $\mathbb{B}$. Adjoining the primitive element of $\mathbb{F'}$ to $\mathbb{E}$, we get $\mathbb{F}=GF(q^{\ell}),\ell=a\ell'$. The new code is defined in $\mathbb{F}$.  We extend the evaluation points to be $A=\{\alpha_1 \mathbb{E}^\ast,\alpha_2 \mathbb{E}^\ast,\dots,\alpha_{n'} \mathbb{E}^\ast\}. $
Since $\{\alpha_1,\alpha_2,...,\alpha_{n'}\}$ are linearly independent over $\mathbb{B}$, we can apply Lemma \ref{ranks for extended fields} and the cosets are distinct.
So, $n=|A|=(q^a-1)n'$.

Assuming the failed node is $f(\alpha^\ast)$ and $\alpha^\ast \in \alpha_i \mathbb{E}^\ast$ and the helpers are chosen from the set $ {R}$, $|R|=d$, the base field for repair is $\mathbb{F}_i$, which is defined by $\mathbb{F}_i\triangleq \mathbb{F}_{q}(\alpha_j, j\in[n+1], j\neq i)$, for $i\in[n]$. 
We define the repair polynomials $\{\eta_t p_j(x), t \in [q_i], j \in [s]\}$, where 
\begin{align}\label{mulitiple_polynomial}
p_{j}(x)=x^{j-1}g(x), j\in[s],x\in\mathbb{F}, 
\end{align}
\begin{align}
g(x)=\prod_{\alpha \in A/ (R\cup\{\alpha^\ast\} ) } (x-\alpha), x\in\mathbb{F},
\end{align}
and $\eta_t$ is the same as that in the original code.
Then, we repair the failed node by
\begin{align}\label{trace equation3}
&tr_{\mathbb{F}/\mathbb{F}_i}(\upsilon_{\alpha^{\ast}}\eta_tp_{j}(\alpha^{\ast})f(\alpha^{\ast}))=-\sum\limits_{\alpha \in A, \alpha \neq\alpha^{\ast}} tr_{\mathbb{F}/\mathbb{F}_i}(\upsilon_{\alpha}\eta_tp_{j}(\alpha)f(\alpha)).
\end{align}

For $x\in \alpha \mathbb{E}^\ast,\alpha\in A'$, we have
\begin{align}
p_{j}(x)=\gamma^{j-1}\alpha^{j-1}g(x), j\in[s],
\end{align}
for some $\gamma \in \mathbb{E}^\ast$. 
If $x \notin R\cup\{\alpha^\ast\}$, since $g(x)=0$, no information is transmitted from node $x$. Next, we consider all other nodes.

For $x=\alpha\gamma,\alpha\in A'$, since $g(x)$ is a constant independent of $j$, $\gamma \in \mathbb{E}\subseteq\mathbb{F}_i$ and $\eta_t,\alpha_i\in\mathbb{F'}$, from Lemma \ref{ranks for extended fields} we have 
\begin{align}
&rank_{\mathbb{F}_i}(\{\eta_tp_{1}(x),\eta_tp_{2}(x),\dots,\eta_tp_{s}(x):t\in[q_i] \})\nonumber\\
=&rank_{\mathbb{F}_i}(\{\eta_t,\eta_t\gamma\alpha,\dots,\eta_t\gamma^{s-1}\alpha^{s-1}:t\in[q_i] \}) \nonumber\\
=&rank_{\mathbb{F}_i}(\{\eta_t,\eta_t\alpha,\dots,\eta_t\alpha^{s-1}:t\in[q_i] \}) \nonumber\\
=&rank_{\mathbb{F'}_i}(\{\eta_t,\eta_t\alpha,\dots,\eta_t\alpha^{s-1}:t\in[q_i]\})\nonumber\\
=&rank_{\mathbb{F'}_i}(\{\eta_tp'_{1}(\alpha),\eta_tp'_{2}(\alpha),\dots,\eta_tp'_{s}(\alpha):t\in[q_i]\}),
\end{align}
which satisfies \eqref{new repair condition}.

When $k\leq d<n-1$, assuming the helpers are randomly chosen from all the remaining nodes, the average repair bandwidth for different choices of the helpers can be calculated as 
\begin{align}
b&=d \left[ \frac{b'a}{d'}\cdot\frac{n-1-(q^a-2)}{n-1}+\ell' a \cdot \frac{q^a-2}{n-1} \right] \label{eq65} \\
&= \frac{d\ell}{d-k+1}+ \frac{d}{n-1}\frac{\ell}{d-k+1}(d-k)(q^a-2). \label{eq66}
\end{align}
Here in \eqref{eq65} the second term corresponds to the helpers in the failed node coset, the first term corresponds to the helpers in the other cosets, and in \eqref{eq66} we used $d'-k'=d-k$.
\end{IEEEproof}

In the case of $d=n-1$, the repair bandwidth of the code in Theorem \ref{second realization of multiple cosets} can be directly calculated from \eqref{new repair bandwidth} as
\begin{align}
b&=ab'(q^a-1)+(q^a-2)\ell \nonumber\\
&=\frac{\ell}{r}(n-1)+\frac{\ell}{r}(r-1)(q^a-2)]. \label{eq77}
\end{align}
In \eqref{eq66} and \eqref{eq77}, the second term is the extra repair bandwidth compared to the original code.

In Theorems \ref{thm:2nd multiple cosets} and \ref{second realization of multiple cosets}, we constructed our schemes by extending previous schemes. However, it should be noted that since we only used the properties of the polynomials $p'_j(x)$, we have no restrictions on the dimensions $k'$ of the original codes. So, in some special cases, even if $k'$ is negative and the original codes do not exist, our theorems still hold. Thus, we can provide more feasible points of $(n,k)$ using our schemes. This is illustrated in the example below.

\begin{exa}
Let us take the $RS(12,8)$ code as an example. We set $q=2,s=4,q_1=5,q_2=9,q_3=13$ and $a=7$. Then, $\ell'=2340$ and $\ell=16380$. Assuming the failed node is $f(\alpha^\ast)$ and $\alpha^\ast\in\alpha_1 C$, then we repair it in $\mathbb{F}_1$ and set the polynomials in \eqref{mulitiple_polynomial}. We can easily check that when $x\in\alpha_1C$, $rank_{\mathbb{F}_1}(\{\eta_tp_{1}(x),\eta_tp_{2}(x),\dots,\eta_tp_{s}(x):t\in[5]\})=20$ and when $x$ in other cosets, $rank_{\mathbb{F}_1}(\{\eta_tp_{1}(x),$ $\eta_tp_{2}(x),\dots,\eta_tp_{s}(x):t\in[5]\})=5$. Therefore, we transmit $100$ symbols in $\mathbb{F}_1$, which can be normalized to $\frac{b}{(n-1)\ell}=0.4545$. Compared with the scheme in \cite{tamo2017optimal}, which need $\ell=2.4\times 10^{19}$ and $\frac{b}{(n-1)\ell}=0.25$, we provide a tradeoff between $\ell$ and $b$.

It should be noted that in this example, the $RS(12,8)$ code needs to be extended from an $RS(3,-1)$ code, which does not exist. However, since we only used the properties of the polynomials $p'_j(x)$ and $p_j(x)$, the new $RS(12,8)$ code still works.
\end{exa}

\subsection{Numerical evaluations and discussions}
\label{sec:single_numerical}

\begin{figure}[!h]
\centering
\includegraphics[width=4in]{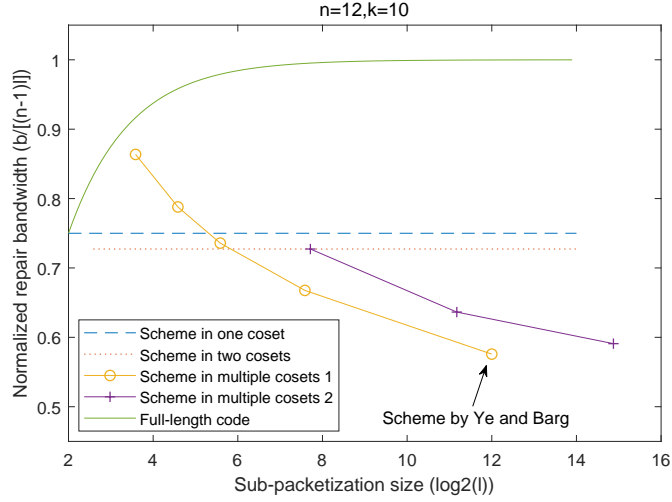}
\caption{Comparison of 3 schemes, $q=2, n=12,k=10,r=2$, x-axis is the log scale sub-packetization size, y-axis is the normalized repair bandwidth.}\label{Comparison of 3 schemes 1}
\end{figure}

\begin{figure}[!h]
\centering
\includegraphics[width=4in]{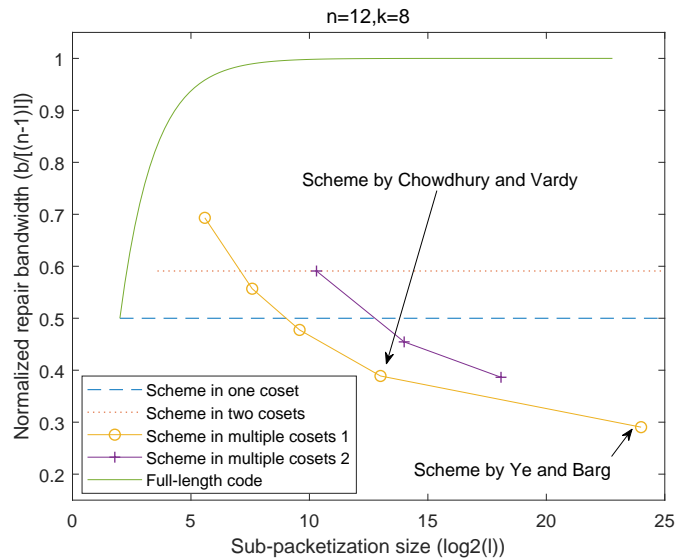}
\caption{Comparison of 3 schemes, $q=2, n=12,k=8,r=4$, x-axis is the log scale sub-packetization size, y-axis is the normalized repair bandwidth. The scheme by Chowdhury and Vardy is in \cite{chowdhury2017improved}, the scheme by Ye and Barg is in \cite{ye2016explicit}, and the full-length code is in \cite{guruswami2017repairing} and \cite{dau2017optimal}.}\label{Comparison of 3 schemes 2}
\end{figure}

In this subsection, we compare the existing and the proposed schemes.  Table \ref{table of comparision1} shows the repair bandwidth and the code length of each scheme. For the comparison, we first show in Figures \ref{Comparison of 3 schemes 1} and \ref{Comparison of 3 schemes 2} the performance of each scheme when the sub-packetization changes, given $(n,k)=(12,10)$ and $(12,8)$, respectively. We only consider $n-1$ helpers. Two single points $(\text{log}_2(\ell)=53.5,\frac{b}{(n-1)\ell}=0.50)$ in $RS(12,10)$ codes and $(\text{log}_2(\ell)=64.4,\frac{b}{(n-1)\ell}=0.25)$ in $RS(12,8)$ codes are not shown in the figures, they can be achieved by both our second realization in multiple cosets and \cite{tamo2017optimal}. We make the following observations.

\begin{enumerate}
    \item For a fixed $(n,k)$, we compare the normalized repair bandwidth $b/[(n-1)\ell]$ in different sub-packetization sizes. In our schemes in multiple cosets, we have a parameter $a$ to adjust the sub-packetization size. From Theorems \ref{thm:2nd multiple cosets} and \ref{second realization of multiple cosets} we know that for the two schemes, $\ell=a\cdot r^{\frac{n}{q^a-1}}$ and $\ell\approx a\cdot(\frac{n}{q^a-1})^{(\frac{n}{q^a-1})}$, respectively, which means that increasing $a$ will decrease the sub-packetization $\ell$. In our schemes in one coset and two cosets, the parameter $a$ is determined by code length $n$, and will not be changed by increasing $\ell$, neither will the normalized repair bandwidth. When $q=2$, $a=1$, the two schemes in multiple cosets coincides with \cite{ye2016explicit} and \cite{tamo2017optimal}, respectively.
     \item The scheme in \cite{chowdhury2017improved} also achieves one tradeoff point in Figure \ref{Comparison of 3 schemes 2}, which can be viewed as a special case of our scheme in multiple coset 1.
    \item For fixed $n,k$, our schemes are better than the full-length code in \cite{guruswami2017repairing} and \cite{dau2017optimal} for all $\ell$, except when $\ell=4$, for which our scheme in one coset is identical to the full-length code. 
   \item While the repair bandwidth of the full-length code grows with $\ell$, our schemes in one coset and two cosets have a constant normalized bandwidth, and our schemes in multiple cosets have a decreasing normalized bandwidth with $\ell$.
    \item For small $\ell$: the schemes in one coset and two cosets are better than those in multiple cosets; when $n=12,k=10, 4\leq \ell \leq 48$, the scheme in two cosets provides the lowest bandwidth; when $n=12,k=8, 4 \leq \ell \leq 768$, one can show that the scheme in one coset has the smallest bandwidth. 
    \item For large $\ell$:  the first realization in multiple cosets has better performance than the second realization in multiple cosets, but our second realization works for any number of helpers. 
\end{enumerate}

\section{Reed-Solomon repair schemes for multiple erasures}
\label{sec:multiple erasures}
In this section, we first present two definitions of the repair schemes for multiple erasures in a MDS code: linear repair scheme definition and dual code repair definition. We prove the equivalence of the two definitions. Then, we present two schemes for repairing multiple erasures in Reed-Solomon codes, where the evaluation points are in one coset and multiple cosets, respectively.

\subsection{Definitions of the multiple-erasure repair}
Let us assume a scalar MDS code $\mathscr{E}$ over $\mathbb{F}=GF(q^\ell)$ has dimension $k$ and code length $n$. Let a codeword be $(C_1,C_2,...C_n)$. Without loss of generality, we assume $\{C_1,C_2,...,C_e\}$ are failed, $e\leq n-k$, and we repair them in the base field $\mathbb{B}=GF(q)$, where $q$ can be any power of a prime number. We also assume that we use all the remaining $d=n-e$ nodes as helpers. The following definitions are inspired by [5] for single erasure.

\begin{defn}\label{linear repair scheme definition}
A linear exact repair scheme for multiple erasures consists of the following.
\begin{enumerate}
    \item A set of queries $Q_{t} \subseteq \mathbb{F}$ for each helper $C_t, e+1\leq t\leq n$. The helper $C_t$ replies with $\{\gamma C_t, \gamma\in Q_{t}\}$.

    \item For each failed node $C_i,i\in[e]$, a linear repair scheme that computes
\begin{align} \label{eq:66}
C_i=\sum_{m=1}^\ell \lambda_{im}\mu_{im},
\end{align}
where $\{\mu_{i1},\mu_{i2},...,\mu_{i\ell}\}$ is a basis for $\mathbb{F}$ over $\mathbb{B}$ and coefficients $\lambda_{im}\in\mathbb{B}$ are $\mathbb{B}$-linear combinations of the replies
\begin{align} \label{eq:67}
\lambda_{im}=\sum_{t=e+1}^n \sum_{\gamma\in Q_{t}}\beta_{im\gamma t}\cdot tr_{\mathbb{F}/\mathbb{B}}(\gamma C_t),
\end{align}
with the coefficients $\beta_{im\gamma t}\in \mathbb{B}$. The repair bandwidth is
\begin{align}
b=\sum_{t=e+1}^\ell rank_{\mathbb{B}}(Q_{t}).
\end{align}
\end{enumerate}
\end{defn}
In the following definition, we consider $e\ell$ dual codewords of $\mathscr{E}$, and index them by $i \in [e], j \in [\ell]$, denoted as $(C'_{ij1},C'_{ij2},\dots,C'_{ijn})$. Since they are dual codwords, we know that $\sum_{t=1}^{n} C_t C'_{ijt}=0$.

\begin{defn}\label{dual code repair definition}
A dual code scheme uses a set of dual codewords $\{(C'_{ij1},C'_{ij2},\dots,C'_{ijn}): i\in[e],j\in[\ell]\}$ that satisfies:
\begin{enumerate}
    \item The {\bf full rank condition}: Vectors
\begin{align}
V_{ij}=(C'_{ij1},C'_{ij2},...,C'_{ije}),i\in[e],j\in[\ell],
\end{align}
are linearly independent over $\mathbb{B}$.
   \item The {\bf repair bandwidth condition}: 
\begin{align}
b=\sum_{t=e+1}^n rank_\mathbb{B}(\{C'_{ijt}:i\in[e],j\in[\ell]\}).
\end{align}
\end{enumerate}

We repair nodes $[e]$ from the linearly independent equations
\begin{align}
\sum_{v=1}^e tr_{\mathbb{F}/\mathbb{B}} (C'_{ijv} C_v) = - \sum_{t=e+1}^n tr_{\mathbb{F}/\mathbb{B}} (C'_{ijt} C_t), i\in[e],j\in[\ell].
\end{align}

\end{defn}
Here we use the same condition names as the single erasure case, but in this section, they are defined for multiple erasures.

\begin{thm}\label{equivalence}
Definitions \ref{linear repair scheme definition} and \ref{dual code repair definition} are equivalent. 
\end{thm}
The equivalence of Definitions \ref{linear repair scheme definition} and \ref{dual code repair definition} follows similarly as arguments in \cite{guruswami2017repairing}, except that we need to first solve $e$ failed nodes simultaneously and then find out the form of each individual failure \eqref{eq:66}. The detailed proof of Theorem \ref{equivalence} is shown in Appendix \ref{sec: appendix equivalence}, part of which uses Lemma \ref{full rand condition for multiple erasures} in Section IV-B.

\begin{rem}
In this paper, we focus on repairing RS code and apply Theorem \ref{equivalence} to RS code. From \cite[Thm. 4 in Ch. 10]{macwilliams1977theory} we know that with the polynomial $p_{ij}(x) \in \mathbb{F}[x]$ for which the degrees are smaller than $n-k$, $(\upsilon_{1}p_{ij}(\alpha_{1}),\upsilon_{2}p_{ij}(\alpha_{2}),\dots,\upsilon_{n}p_{ij}(\alpha_{n}))$ is the dual codeword of $RS(n,k)$, where $\upsilon_{i}, i\in[n]$ are non-zero constants determined by the evaluation points set $A$. So, in RS code, Definition \ref{dual code repair definition} reduces to finding polynomials $p_{ij}(x)$ with degrees smaller than $n-k$. In what follows we use $p_{ij}(\alpha_t)$ to replace the dual codeword symbol $C'_{ijt}$ in Definition \ref{dual code repair definition} for RS code. One can easily show that the constants $\upsilon_i, i \in [n]$ do not affect the ranks in the full rank condition and the repair bandwidth condition.
\end{rem}

\subsection{Multiple-erasure repair in one coset}
There are several studies about the multiple erasures for full-length RS codes \cite{dau2018repairing} and \cite{mardia2018repairing}. Inspired by these works, we propose our scheme for multiple erasures in one coset.

From Theorem \ref{equivalence}, we know that finding the repair scheme for multiple erasures in RS code is equivalent to finding dual codewords (or polynomials) that satisfy the full rank condition and repair bandwidth condition. Given a basis $\{\xi_1,\xi_2,...,\xi_\ell \}$ for $\mathbb{F}$ over $\mathbb{B}$, we define some matrices as below. They are used to help us check the two rank conditions according to Lemmas \ref{full rand condition for multiple erasures} and \ref{matrix rank}, whose proofs are shown in Appendices C and D, respectively. Let the evaluation points of an RS code over $\mathbb{F}$ be $A=\{\alpha_1,\dots,\alpha_n\}$. Let $p_{ij}(x), i \in [e], j \in [\ell],$ be polynomials over $\mathbb{F}$, and $\mathbb{B}$ a subfield of $\mathbb{F}$.  Define
\begin{align}\label{repair matrix form}
S_{it}=\begin{bmatrix}tr_{\mathbb{F}/\mathbb{B}}(\xi_{1}p_{i1}(\alpha_t))&\cdots\ &tr_{\mathbb{F}/\mathbb{B}}(\xi_{\ell}p_{i1}(\alpha_t))\\tr_{\mathbb{F}/\mathbb{B}}(\xi_{1}p_{i2}(\alpha_t))&\cdots\ &tr_{\mathbb{F}/\mathbb{B}}(\xi_{\ell}p_{i2}(\alpha_t))\\ \vdots  & \ddots  & \vdots  \\ tr_{\mathbb{F}/\mathbb{B}}(\xi_{1}p_{i\ell}(\alpha_t))&\cdots\ &tr_{\mathbb{F}/\mathbb{B}}(\xi_{\ell}p_{i\ell}(\alpha_t))\end{bmatrix},
\end{align}
\begin{align}\label{combined matrix}
S \triangleq \begin{bmatrix}S_{11}&S_{12}&\cdots\ &S_{1e}\\S_{21}&S_{22}&\cdots\ &S_{2e}\\ \vdots  & \vdots  & \ddots  & \vdots  \\ S_{e1}&S_{e2}&\cdots\ &S_{ee}\end{bmatrix}.
\end{align}

\begin{lem}\label{full rand condition for multiple erasures}
The following two statements are equivalent:
\begin{enumerate}
    \item Vectors $V_{ij}=(p_{ij}(\alpha_1),p_{ij}(\alpha_2),\dots,p_{ij}(\alpha_e)),i\in[e],j\in[\ell]$ are linearly independent over $\mathbb{B}$.

    \item Matrix $S$ in \eqref{combined matrix} has full rank.

\end{enumerate}
\end{lem}

\begin{lem}\label{matrix rank}
For $t \in [n]$, consider $S_{it}$ in \eqref{repair matrix form}, 
\begin{align}
rank (\begin{bmatrix}S_{1t}\\S_{2t}\\\vdots\\S_{et}\end{bmatrix})=rank_\mathbb{B}(\{p_{ij}(\alpha_t):i\in[e],j\in[\ell] \}).
\end{align}
\end{lem}

\begin{thm}
Let $q$ be a prime number.
There exists an $RS(n,k)$ code over $\mathbb{F}=GF(q^\ell)$ of which $n<q^a, q^s\leq r$ and $a|\ell$, such that the repair bandwidth for $e$ erasures is $b\leq\frac{e\ell}{a}(n-e)(a-s)$  measured in symbols over $\mathbb{B}$, for $e$ satisfying $a\geq \frac{e(e-1)}{2}(a-s)^2$.
\end{thm}

\begin{IEEEproof}
We define the code over the field $\mathbb{F}=GF(q^\ell)$ extended by $\mathbb{E}=GF(q^a)$, where $\beta$ is the primitive element of $\mathbb{F}$. The evaluation points are chosen to be $A=\{\alpha_{1},\alpha_{2},\dots,\alpha_{n}\} \subseteq \mathbb{E}^\ast$, which is one of the cosets in Lemma \ref{cyclic subgroups}. Without loss of generality, we assume the $e$ failed nodes are $\{\alpha_1,\alpha_2,\dots,\alpha_e\}$. The base field is $\mathbb{B}=GF(q)$.

\noindent\textbf{Construction III}:
We first consider the special case when $s=a-1$. In this case, inspired by \cite[Proposition 1]{ mardia2018repairing}, we choose the polynomials 
\begin{equation}
p_{ij}(x)=\frac{\delta_{i}tr_{\mathbb{E}/\mathbb{B}}(\frac{\mu_{j}}{\delta_{i}}(x-\alpha_i))}{x-\alpha_i}, i\in[e],j\in[a],
\end{equation}
where $\{\mu_{1},\mu_{2},\dots,\mu_{a}\}$ is the basis for $\mathbb{E}$ over $\mathbb{B}$, and $\delta_i \in \mathbb{E}, i\in [e],$ are coefficients to be determined. From \cite[Theorem 3]{ mardia2018repairing}, we know that for $a>\frac{e(e-1)}{2}$,  there exists $\delta_i,i\in[e]$ such that $p_{ij}(x)$ satisfy the full rank condition: the vectors $V_{ij}=(p_{ij}(\alpha_1),p_{ij}(\alpha_2),\dots,p_{ij}(\alpha_e)),i\in[e],j\in[a]$ are linearly independent over $\mathbb{B}$ and the repair bandwidth condition:
\begin{align}\label{rank of multiple erasures1}
\sum_{t=e+1}^n rank_{\mathbb{B}}(\{p_{ij}(\alpha_t):i\in[e],j\in[a] \})=(n-e)e-\frac{e(e-1)(q-1)}{2}.
\end{align}

 Then, let \{$\eta_1, \eta_2, \dots,\eta_{\ell/a}\}$ be a set of basis for $\mathbb{F}$ over $\mathbb{E}$; we have the $e\ell$ polynomials as  $\{\eta_w p_{ij}(x): w\in[\ell/a], i\in [e],j\in [a]\}$. Since \{$\eta_1, \eta_2, \dots,\eta_{\ell/a}\}$ are linearly independent over $\mathbb{E}$ and for any $b_{ijw}\in\mathbb{B},b_{ijw}p_{ij}(x)\in\mathbb{E}$, we have 
\begin{align}
\sum_{i,j,w}b_{ijw}\eta_wV_{ij}=0\iff \sum_{i,j}b_{ijw}V_{ij}=0,\forall w\in[\frac{\ell}{a}].
\end{align}
Also, we know that there does not exist nonzero $b_{ijw}\in\mathbb{B}$ that satisfies $\sum_{i,j}b_{ijw}V_{ij}=0$, so we have that vectors $\{\eta_wV_{ij}$, $w\in[\ell/a],i\in[e],j\in[a]\}$ are also linearly independent over $\mathbb{B}$. So, from Definition \ref{dual code repair definition}, we know that we can recover the failed nodes and the repair bandwidth is
\begin{align}
b=&rank_{\mathbb{B}}(\{\eta_1 p_{ij}(x), \eta_2 p_{ij}(x),\dots, \eta_{\ell/a}p_{ij}(x):i\in[e],j\in[a] \})\nonumber\\
=&\frac{\ell}{a}rank_{\mathbb{B}}(\{p_{ij}(x),i\in[e],j\in[a] \})\nonumber\\
=&\frac{\ell}{a}\left[(n-e)e-\frac{e(e-1)(q-1)}{2}\right].
\end{align}

\noindent\textbf{Construction IV}:
For $s\leq a-1$, consider the polynomials
\begin{equation}
p_{ij}(x)=\delta_{i}^{q^s-1}\mu_{j}\prod\limits_{\varepsilon=1}^{q^s-1}\left(x-\left(\alpha_i-w_\varepsilon^{-1}\frac{\mu_{j}}{\delta_i}\right)\right), j\in[a],
\end{equation}
where $\{\mu_{1},\mu_{2},\dots,\mu_{a}\}$ is the basis for $\mathbb{E}$ over $\mathbb{B}$, $W=\{w_0 = 0, w_{1},w_{2},\dots,w_{q^s-1}\}$ is an $s$-dimensional subspace in $\mathbb{E}$, $s<a,q^s\leq r$, and $\delta_i \in \mathbb{E}, i\in [e],$ are coefficients to be determined.

When $x=\alpha_i$, we have
\begin{align}
p_{ij}(\alpha_i)=\mu_{j}^{q^s}\prod\limits_{\varepsilon=1}^{q^s-1}w_\varepsilon^{-1}.
\end{align}
Since $\prod\limits_{\varepsilon=1}^{q^s-1}w_\varepsilon^{-1}$ is a constant, from Lemma \ref{basis lemma1} we have
\begin{align}
rank_{\mathbb{B}}(\{p_{i1}(\alpha_i),p_{i2}(\alpha_i),\dots,p_{ia}(\alpha_i) \})=a.
\end{align}

For $x\neq \alpha_i$, set $x'=\alpha_i-x$, we have
\begin{align}\label{polynomials simplify construction 2}
p_{ij}(x)&=\delta_{i}^{q^s-1}\mu_{j}\prod\limits_{\varepsilon=1}^{q^s-1}\left(w_\varepsilon^{-1}\frac{\mu_{j}}{\delta_i}-x'\right)\nonumber\\
&=\delta_{i}^{q^s-1}\mu_{j}\prod\limits_{\varepsilon=1}^{q^s-1}(w_\varepsilon^{-1}x')\prod\limits_{\varepsilon=1}^{q^s-1}\left(\frac{\mu_{j}}{\delta_i x'}-w_\varepsilon\right)\nonumber\\
&=(\delta_ix')^{q^s}\prod\limits_{\varepsilon=1}^{q^s-1}(w_\varepsilon^{-1})\prod\limits_{\varepsilon=0}^{q^s-1}\left(\frac{\mu_{j}}{\delta_i x'}-w_\varepsilon\right).
\end{align}
By \cite[p. 4]{goss2012basic}, $g(y)=\prod\limits_{\varepsilon=0}^{q^s-1}(y-w_\varepsilon)$ is a linear mapping from $\mathbb{E}$ to itself with dimension $a-s$ over $\mathbb{B}$. Since $(\delta_ix')^{q^s}\prod\limits_{\varepsilon=1}^{q^s-1}(w_\varepsilon^{-1})$ is a constant independent of $j$, we have
\begin{align}\label{multiple erasures in one coset polynomials rank}
rank_{\mathbb{B}}(\{p_{i1}(x),p_{i2}(x),\dots,p_{ia}(x) \})\leq a-s,
\end{align}
which means that $p_{ij}(x)$ can be written as
\begin{align}
p_{ij}(x)=\delta_i^{q^s}\sum_{v=1}^{a-s}\rho_{jv}\lambda_v,
\end{align}
where $\{\lambda_1, \lambda_2,...,\lambda_{a-s}\}$ are linearly independent over $\mathbb{B}$, $\rho_{jv}\in\mathbb{B}$, and they are determined by $\delta_i,\mu_j$ and $x-\alpha_i$.

From Lemma \ref{full rand condition for multiple erasures}, we know that if the matrix $S$ in \eqref{combined matrix} has full rank, then we can recover the $e$ erasures. It is difficult to directly discuss the rank of the matrix, but assume that the polynomials above satisfy the following two conditions:
\begin{enumerate}
    \item $S_{ii},i\in [e]$ are identity matrices.

    \item For any fixed $i \in [e]$,
    \begin{align}\label{matrix elimination}
S_{it}\cdot S_{ty}=\pmb{0}_{\ell \times \ell}, i>t,y>t.
\end{align}
\end{enumerate}
Then, it is easy to see that through Gaussian elimination, we can transform the matrix $S^T$ to an upper triangular block matrix, which has identity matrices in the diagonal. Hence, $S$ has full rank. 

Here, we choose $\{\xi_1,\xi_2,...,\xi_\ell\}$ to be the dual basis of $\{\mu_{1}^{q^s}\prod\limits_{\varepsilon=1}^{q^s-1}w_\varepsilon^{-1},\mu_{2}^{q^s}\prod\limits_{\varepsilon=1}^{q^s-1}w_\varepsilon^{-1},...,\mu_{\ell}^{q^s}\prod\limits_{\varepsilon=1}^{q^s-1}w_\varepsilon^{-1}\}$, so 
\begin{equation}
tr_{\mathbb{F}/\mathbb{B}}(\xi_{m}p_{ij}(\alpha_i))=
\begin{cases}
0 , m\neq j,\\
1 , m=j.
\end{cases}
\end{equation}
Therefore, $S_{ii},i\in [e]$ are identity matrices. 
We set $\delta_1=1$, and recursively choose $\delta_i$ after choosing $\{\delta_1, \delta_2,..., \delta_{i-1}\}$ to satisfy \eqref{matrix elimination}. Define $\delta_i'=\delta_i^{q^s}$, and $c_{mp}$ to be the $(m,p)$-th element in $S_{ty}$ for $m,p\in[a]$. \eqref{matrix elimination} can be written as
\begin{align}\label{matrix elimination construction 2}
\sum_{m=1}^a c_{mp} tr_{\mathbb{F}/\mathbb{B}}(\xi_m p_{ij}(\alpha_t)) =
\sum_{m=1}^a c_{mp}\sum_{v=1}^{a-s}b_{jv}tr_{\mathbb{F}/\mathbb{B}}(\xi_m\delta'_i \lambda_v)=0,
\forall j\in[a],
\end{align}
where $\lambda_v, v\in [a-s],$ are determined by $\delta_i,\mu_j$ and $\alpha_t-\alpha_i$. Equation \eqref{matrix elimination construction 2} is satisfied if 
\begin{align}\label{matrix elimination construction 2 simplify}
\sum\limits_{m=1}^a c_{mp}tr_{\mathbb{F}/\mathbb{B}}(\xi_m\delta'_i \lambda_v)=0,v\in[a-s],p\in[a].
\end{align}
As a special case of Lemma \ref{matrix rank}, we have
\begin{align}
rank (S_{ty})=rank_\mathbb{B}(\{p_{tj}(\alpha_y),j\in[\ell] \}).
\end{align}
Then, from \eqref{multiple erasures in one coset polynomials rank} we know that the rank of $S_{ty}$ is at most $a-s$, which means in \eqref{matrix elimination construction 2 simplify} we only need to consider $p$ corresponding to the independent $a-s$ columns of $S_{ty}$. So, \eqref{matrix elimination construction 2 simplify} is equivalent to $(a-s)^2$ linear requirements. For $\delta'_i \in \mathbb{E}$, we can view it as $a$ unknowns over $\mathbb{B}$, and we have
\begin{align}
\frac{(2e-i)(i-1)}{2}(a-s)^2\le \frac{e(e-1)}{2}(a-s)^2
\end{align}
linear requirements over $\mathbb{B}$ according to \eqref{matrix elimination}. Also knowing $\delta'_i$, we can solve $\delta_i={\delta_i}^{q^\ell}={{\delta'}_i}^{q^{\ell-s}}$. Therefore, we can find appropriate $\{\delta_1,\delta_2,\dots,\delta_e\}$ to make matrix $S$ full rank when 
\begin{align}
a\geq \frac{e(e-1)}{2}(a-s)^2.
\end{align}

Then, let \{$\eta_1, \eta_2, \dots,\eta_{\ell/a}\}$ be a basis for $\mathbb{F}$ over $\mathbb{E}$, we have the $e\ell$ polynomials as  $\{\eta_w p_{ij}(x), w\in[\ell/a], i\in [e],j\in [a]\}$. Similar to Construction III, we know that vectors $\{\eta_wV_{ij}$, $w\in[\ell/a],i\in[e],j\in[a]\}$ are linearly independent over $\mathbb{B}$. Therefore, we can recover the failed nodes and the repair bandwidth is
\begin{align}\label{repair bandwidth of multiple erasures one coset}
b=&rank_{\mathbb{B}}(\{\eta_1 p_{ij}(x), \eta_2 p_{ij}(x),\dots, \eta_{\ell/a}p_{ij}(x):i\in[e],j\in[a] \})\nonumber\\
=&\frac{\ell}{a}rank_{\mathbb{B}}(\{p_{ij}(x):i\in[e],j\in[a] \})\nonumber\\
\leq&\frac{e\ell}{a}(n-e)(a-s).
\end{align}
Thus, the proof is completed. 
\end{IEEEproof}

In our scheme, we have constructions for arbitrary $a, s$, such that $a \mid \ell, s \le a-1$, while the existing schemes in \cite{dau2018repairing} and \cite{mardia2018repairing} mainly considered the special case $\ell=a$. It should be noted that the scheme in \cite{mardia2018repairing} can also be used in the case of $s=a-1$ over $\mathbb{E}$ with repair bandwidth $(n-e)e-\frac{e(e-1)(q-1)}{2}$. And, with $\ell/a$ copies of the code, it can also reach the same repair bandwidth of our scheme. However, by doing so, the code is a vector code, but our scheme constructs a scalar code.

\subsection{Multiple-erasure repair in multiple cosets} 
Recall that the scheme in Theorem \ref{second realization of multiple cosets} for a single erasure is a small sub-packetization code with small repair bandwidth for any number of helpers. When there are $e$ erasures and $d$ helpers, $e\le n-k, k \le d \le n-e$, we can recover the erasures one by one using the $d$ helpers. However, inspired by \cite{ye2017repairing}, the repaired nodes can be viewed as additional helpers and thus we can reduce the total repair bandwidth. Finally, for every helper, the transmitted information for different failed nodes has some overlap, resulting in a further bandwidth reduction.

The approach we take is similar to that of Section \ref{sec: 1 erasure multi cosets}. We take an original code and extend it to a new code with evaluation points as in \eqref{extended evaluation points}. If a helper is in the same coset as any failed node, it transmits naively its entire data; otherwise, it transmits the same amount as the scheme in the original code. After the extension, the new construction decreases the sub-packetization size for fixed $n$, and the bandwidth is only slightly larger than the original code.

\begin{figure}[!h]
\centering
\includegraphics[width=3.2in]{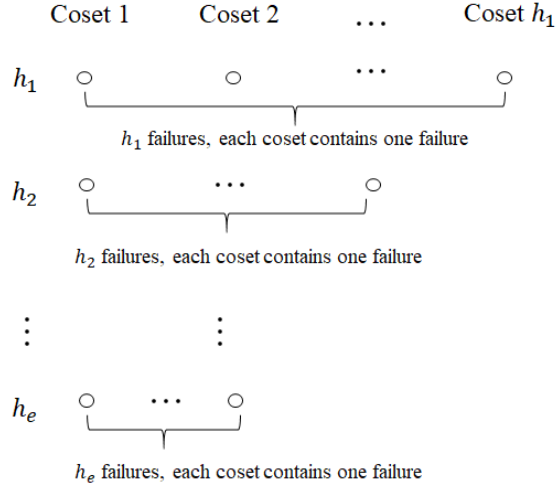}
\caption{Location of the erasures. $e$ erasures are located in $h_1$ cosets. For $i\in[e]$, we set $0\leq h_i\leq e, h_1\geq h_2 \geq...\geq h_e$ and $\sum_{i=1}^eh_i=e$.}\label{location of erasures}
\end{figure}

The location of the $e$ erasures are described by $h_i, i\in[e],$ where $0\leq h_i\leq e, h_1\geq h_2 \geq...\geq h_e$, $\sum_{i=1}^eh_i=e$. We assume the erasures are located in $h_1$ cosets, and after removing one erasure in each coset, the remaining erasures are located in $h_2$ cosets. Then, for the remaining erasures, removing one in each coset, we get the rest of erasures in $h_3$ cosets, and so on. Figure \ref{location of erasures} also shows the erasure locations described above.

In our scheme, we first repair $h_1$ failures, one from each of the $h_1$ cosets. Then, for $2 \le i \le e$, we repeat the following:  After repairing $h_1,h_2,...,h_{i-1}$ failures, we view these repaired nodes as helpers and repair next $h_i$ failures, one from each of the $h_i$ cosets.

The repair bandwidth of the scheme is showed in the following theorem.

\begin{thm}\label{theorem for multiple erasures in multiple cosets}
Let $q$ be a prime number. There exists an $RS(n,k)$ code over $\mathbb{F}=GF(q^{\ell})$ for which $\ell=asq_1q_2...q_{\frac{n}{q^a-1}}$, where $q_i$ is the  $i$-th prime number that satisfies $s|(q_i-1), s=(n-k)!$ and $a$ is an integer. For $e$ erasures and $d$ helpers, $e \le n-k, k\le d \le n-e$, the average repair bandwidth measured in symbols over $\mathbb{B}$ is
\begin{align}\label{average_repair_bandwidth_for_multiple_erasures_in_multiple_cosets}
b \le &\frac{d\ell}{(n-e)}\bigg[(h_1(q^a-1)-e) + (n-h_1(q^a-1))\sum_{i=1}^e \frac{h_i}{d-k+\sum_{v=1}^ih_v}\bigg],
\end{align}
where $h_i,i\in[e]$ are the parameters that define the location of erasures in Fig. \ref{location of erasures}.
\end{thm}

\begin{IEEEproof}
We first prove the case when $a$ and $\ell'$ are relatively prime using Lemma \ref{ranks for extended fields}, the case when $a$ and $\ell'$ are not necessarily relatively prime are proved in Appendix A. We use the code in \cite{ye2017repairing} as the original code. Let $\mathbb{F}_{q}(\alpha)$ be the field obtained by adjoining $\alpha$ to the base field $\mathbb{B}=GF(q)$. Similarly let  $\mathbb{F}_{q}(\alpha_1,\alpha_2,\dots,\alpha_n)$ be the field for adjoining multiple elements. Let $\alpha_i$ be an element of order $q_i$ over $\mathbb{B}$ and $h$ be the number of erasures in the original code. The original code is defined in the field $\mathbb{F'}=GF(q^{\ell'})=GF(q^{s q_1 q_2\dots q_{n'}})$, which is the degree-s of extension of $\mathbb{F}_{q}(\alpha_1, \alpha_2, \dots \alpha_{n'})$. The evaluation points are $A'=\{\alpha_1, \alpha_2, \dots \alpha_{n'}\}$. 
The subfield $\mathbb{F'}_{[h]}$ is defined as $\mathbb{F'}_{[h]}=\mathbb{F}_{q}(\alpha_j, j=h+1,h+2,\dots,n')$, and $\mathbb{F'}_i$ is defined as $\mathbb{F}_q(\alpha_j, j \neq i, j \in [n'])$.

In the original code, we assume without loss of generality that there are $h$ failed nodes $f'(\alpha_1)$, $f'(\alpha_2),\dots,f'(\alpha_h)$. Consider the polynomials for failed node $f'(\alpha_i), 1 \le i \le h,$ as
\begin{align}
p'_{ij}(x)=x^{j-1}g'_i(x), j\in[s_i],x\in\mathbb{F'},
\end{align}
where
\begin{align}
g'_i(x)=\prod_{ \alpha \in A'/(R' \cup \{\alpha_i, \alpha_{i+1},\dots\alpha_h\}) }(x-\alpha), 
x\in\mathbb{F'},
\end{align}
for $R' \subseteq A', |R'|=d'$ being the set of helpers. 
The set of repair polynomials are $\{\eta_{it }p'_{ij}(x), i \in [h], j \in [s_i], t \in[\frac{sq_i}{s_i}] \}$, where $\eta_{it }\in \mathbb{F'}$ are constructed in \cite{ye2017repairing} to ensure that $\{\eta_{it }p'_{i1}(\alpha_i),\eta_{it }p'_{i2}(\alpha_i),\dots,\eta_{it }p'_{is_i}(\alpha_i)\}$ forms the basis for $\mathbb{F'}$ over $\mathbb{F'}_i$.

Then, the failed nodes are repaired one by one from
\begin{align}
&tr_{\mathbb{F'}/\mathbb{F'}_{i}}(\upsilon_{\alpha_i}\eta_{it }p'_{ij}(\alpha_{i})f'(\alpha_{i}))=-\sum\limits_{\epsilon=1,\epsilon\neq i}^{n}tr_{\mathbb{F'}/\mathbb{F'}_{i}}(\upsilon_{\epsilon}\eta_{it }p'_{ij}(\alpha_{\epsilon})f'(\alpha_{\epsilon})).
\end{align}
 For $x\notin R'\cup\{\alpha_i, \alpha_{i+1},\dots\alpha_h\}$, $p'_{ij}(x)=0$ and no information is transmitted. Once $f'(\alpha_i)$ is recovered, it is viewed as a new helper for the failures $i+1,i+2,\dots,h$.

Since $\mathbb{F'}_{[h]} \leq \mathbb{F'}_i$, the information transmitted  from the helper $\alpha_\epsilon$ can be represented as
\begin{align}
&tr_{\mathbb{F'}/\mathbb{F'}_{i}}(\upsilon_{\epsilon}\eta_{it }p'_{ij}(\alpha_{\epsilon})f'(\alpha_{\epsilon}))\nonumber\\
=&tr_{\mathbb{F'}/\mathbb{F'}_{i}}\left( \xi'_{im} \sum\limits_{m=1}^{q'_i}  tr_{\mathbb{F'}_i/\mathbb{F'}_{[h]}} ( \upsilon_{\epsilon}\eta_{it } \xi_{im} p'_{ij}(\alpha_{\epsilon})f'(\alpha_{\epsilon})) \right)\nonumber\\
=&\sum\limits_{m=1}^{q'_i}\xi'_{im}tr_{\mathbb{F'}/\mathbb{F'}_{[h]}}(\upsilon_{\epsilon}\eta_{it }\xi_{im}p'_{ij}(\alpha_{\epsilon})f'(\alpha_{\epsilon})), \label{eq:97}
\end{align}
where $q'_i=\frac{q_1q_2\dots q_h}{q_i}$, $\{\xi_{i1},\xi_{i2},\dots,\xi_{iq'_i}\}$ and $\{\xi'_{i1},\xi'_{i2},\dots,\xi'_{iq'_i}\}$ are the dual basis for $\mathbb{F'}_i$ over $\mathbb{F'}_{[h]}$. We used the fact that $tr_{\mathbb{F'}/\mathbb{F'}_{i}}( tr_{\mathbb{F'}_i/\mathbb{F'}_{[h]}} (\cdot)) = tr_{\mathbb{F'}/\mathbb{F'}_{[h]}} (\cdot) $, for 
$\mathbb{F'}_{[h]} \leq \mathbb{F'}_{i} \leq \mathbb{F'}$.

The original code satisfies the full rank condition for every $i\in [h]$, and each helper $\alpha_\epsilon$ transmits \cite{ye2017repairing}
\begin{align} \label{eq:106}
&rank_{\mathbb{F'}_{[h]}}\bigg(\{\eta_{it }\xi_{im}p'_{ij}(\alpha_{\epsilon}): i\in[h], j\in[s_i],t \in[\frac{sq_i}{s_i}],m\in [q'_i]\}\bigg)\nonumber\\
=&rank_{\mathbb{F'}_{[h]}}\bigg(\{\eta_{it }\xi_{im}: i\in[h],t \in[\frac{sq_i}{s_i}],m\in[q'_i]\}\bigg)\nonumber\\
=&\frac{h\ell'}{(d'-k'+h)\prod_{v=h+1}^{n'}p_v}
\end{align}
symbols over $\mathbb{F'}_{[h]}$, which achieves the MSR bound.
 
In our new code, we extend the field to $\mathbb{F}=GF(q^{\ell}),\ell=a\ell',$ by adjoining an order-$a$ element $\alpha_{n+1}$ to $\mathbb{F}$. We set $d-k=d'-k'$. The new evaluation points consist of $A=\{\alpha_1 \mathbb{E}^\ast,\alpha_2 \mathbb{E}^\ast,\dots,\alpha_n' \mathbb{E}^\ast\}, \mathbb{E}=GF(q^a)=\mathbb{F}_{q}(\alpha_{n+1})$. 
The subfield $\mathbb{F}_{[h]}$ is defined by adjoining $\alpha_{n+1}$ to $\mathbb{F'}_{[h]}$, and $\mathbb{F}_i$ is defined as $\mathbb{F}_q(\alpha_j, j \neq i, j \in [n+1])$.

Assume first that each coset contains at most one failure, and there are $h$ failures in total.
We assume without loss of generality that the evaluation points of the $h$ failed nodes are in $\{\alpha_1 \mathbb{E}^\ast,\alpha_2 \mathbb{E}^\ast,\dots,\alpha_h \mathbb{E}^\ast\}$, and they are $\alpha_1\gamma_1,\alpha_2\gamma_2,\dots,\alpha_h\gamma_h$ for some $\gamma_w\in\mathbb{E}, w \in [h].$ Let the set of helpers be $ {R} \subseteq A$, $|R|=d$. We define the polynomials
\begin{align}
p_{ij}(x)=x^{j-1}g_i(x), j\in[s_i],x\in\mathbb{F}, 
\end{align}
where 
\begin{align}
g_i(x)=\prod_{\alpha\in A/\{ R \cup\{\alpha_i\gamma_i, \alpha_{i+1}\gamma_{i+1},\dots\alpha_h\gamma_h\}\}}(x-\alpha), x\in\mathbb{F}.
\end{align}
The set of repair polynomials are $\{\eta_{it }p_{ij}(x), i \in [h], j \in [s_i], t \in[\frac{sq_i}{s_i}] \}$, where $\eta_{it }\in \mathbb{F'}$ are the same as the original construction.
We use field $\mathbb{F}_i$ as the base field for the repair. 
\begin{align}
&tr_{\mathbb{F}/\mathbb{F}_i}(\upsilon_{\alpha_i\gamma_i}\eta_{it}p_{ij}(\alpha_{i}\gamma_i)f(\alpha_{i}\gamma_i))=-\sum\limits_{\alpha \in A, \alpha \neq\alpha_i\gamma_i}tr_{\mathbb{F}/\mathbb{F}_i}(\upsilon_{\alpha}\eta_{it}p_{ij}(\alpha)f(\alpha)).
\end{align}
If $x \in R \cup\{\alpha_i\gamma_i, \alpha_{i+1}\gamma_{i+1},\dots\alpha_h\gamma_h\}$, $p_{ij}(x)=0$ and no information is transmitted. Next, we consider all other nodes.

If $x = \alpha_i \gamma$ for some $\gamma \in \mathbb{E}^\ast$, we have
\begin{align}
p_{ij}(x)=\gamma^{j-1}\alpha_i^{j-1} g_i(x).
\end{align}

Since $\eta_{it},\alpha_i\in\mathbb{F'}$ and $g_i(x)$ is a constant independent of $j$, we have 
\begin{align}\label{multiple erasures in multiple cosets full rank}
&rank_{\mathbb{F}_i}(\{\eta_{it }p_{i1}(x),\eta_{it }p_{i2}(x),\dots,\eta_{it }p_{is_i}(x):t \in[\frac{sq_i}{s_i}]\})\nonumber\\
=&rank_{\mathbb{F}_i}(\{\eta_{it },\eta_{it }\alpha_i,\dots,\eta_{it }\alpha_i^{s-1}:t \in[\frac{sq_i}{s_i}]\})\nonumber\\
=&rank_{\mathbb{F}_i'}(\{\eta_{it }p'_{i1}(\alpha_i),\eta_{it }p'_{i2}(\alpha_i),\dots,\eta_{it }p'_{is_i}(\alpha_i): t \in[\frac{sq_i}{s_i}]\}) 
\end{align}
which indicates the full rank. Note that the last equation follows from Lemma \ref{ranks for extended fields}. As a result we can recover the failed nodes and each helper in the cosets containing the failed nodes transmit $\ell$ symbols in $\mathbb{B}$.

For $x = \alpha_\epsilon \gamma, \epsilon>h,$ since $\mathbb{F}_{[h]}$ is a subfield of $\mathbb{F}_i$ and from Lemma \ref{ranks for extended fields} we know that $\{\xi_{i1},\xi_{i2},\dots,\xi_{iq'_i}\}$ and $\{\xi'_{i1},\xi'_{i2},\dots,\xi'_{iq'_i}\}$ are also the dual basis for $\mathbb{F}_i$ over $\mathbb{F}_{[h]}$, then, similar to \eqref{eq:97}, we have
\begin{align}
&tr_{\mathbb{F}/\mathbb{F}_i}(\upsilon_{\alpha}\eta_{it}p_{ij}(x)f(x))=\sum\limits_{m=1}^{q'_i}\xi'_{im}tr_{\mathbb{F}/\mathbb{F}_{[h]}}(\upsilon_{\epsilon}\eta_{it }\xi_{im}p_{ij}(x)f(x)).
\end{align}
Using the fact that $g_i(x)$ is a constant independent of $j$, $x\in\mathbb{F}_{[h]}$ and $\eta_{it }\xi_{im} \in \mathbb{F'}$, from Lemma \ref{ranks for extended fields} we know that
\begin{align}\label{multiple erasures in multiple cosets repair bandwidth}
&rank_{\mathbb{F}_{[h]}}\bigg(\{\eta_{it }\xi_{im}p_{ij}(x):i\in[h], j\in[s_i],t \in[\frac{sq_i}{s_i}],m\in[q'_i]\}\bigg)\nonumber\\
=&rank_{\mathbb{F}_{[h]}}\bigg(\{\eta_{it }\xi_{im}: i\in[h],t \in[\frac{sq_i}{s_i}],m\in[q'_i]\}\bigg)\nonumber\\
=&rank_{\mathbb{F'}_{[h]}}\bigg(\{\eta_{it }\xi_{im}: i\in[h],t \in[\frac{sq_i}{s_i}],m\in[q'_i]\}\bigg)\nonumber\\
=&rank_{\mathbb{F'}_{[h]}}\bigg(\{\eta_{it }\xi_{im}p'_{ij}(\alpha_{\epsilon}): i\in[h], j\in[s_i],t \in[\frac{sq_i}{s_i}],m\in[q'_i]\}\bigg)\nonumber\\
=&\frac{h\ell'}{(d-k+h)\prod_{v=h+1}^{n'}q_v}, 
\end{align}
where the last equality follows from \eqref{eq:106} and $d'-k'=d-k$.
So, each helper in the other cosets transmits $\frac{h\ell}{d-k+h}$ symbols over $\mathbb{B}$.

Using the above results, we calculate the repair bandwidth in two steps.

\noindent\textbf{Step 1.} We first repair $h_1$ failures, one from each of the $h_1$ cosets. From \eqref{multiple erasures in multiple cosets full rank}, we know that in the $h_1$ cosets containing the failed nodes, we transmit $\ell$ symbols over $\mathbb{B}$. By \eqref{multiple erasures in multiple cosets repair bandwidth}, for each helper in other cosets, we transmit $\frac{h_1\ell}{d-k+h_1}$ symbols over $\mathbb{B}$.

\noindent\textbf{Step 2.} For $2 \le i \le e$, repeat the following. After repairing $h_1,h_2,...,h_{i-1}$ failures, these nodes can be viewed as helpers for repairing next $h_i$ failures, one from each of the $h_i$ cosets. So, we have $d+\sum\limits_{v=1}^{i-1}h_v$ helpers for the $h_i$ failures. For the helpers in the $h_1$ cosets containing the failed nodes, we already transmit $\ell$ symbols over $\mathbb{B}$ in Step 1 and no more information needs to be transmitted. For each helper in other cosets, we transmit $\frac{h_i\ell}{d-k+\sum_{v=1}^i h_v}$ symbols over $\mathbb{B}$.

Thus, we can repair all the failed nodes. The repair bandwidth can be calculated as \eqref{average_repair_bandwidth_for_multiple_erasures_in_multiple_cosets}.
%
%
\end{IEEEproof}

Suppose that $e$ failures are to be recovered. Compared to the naive strategy which always uses $d$ helpers to repair the failures one by one, our scheme gets a smaller repair bandwidth since the recovered failures are viewed as new helpers and we take advantage of the overlapped symbols for repairing different failures similar to \cite{ye2017repairing}. 

In the case when $n\gg e (q^a-1)$, or when we arrange nodes with correlated failures in different cosets, we can assume that all the erasures are in different cosets, $h_1=e, h_2=h_3=...=h_e=0$. For example, if correlated failures tend to appear in the same rack in a data center, we can assign each node in the rack to a different coset. Under such conditions, we simplify the repair bandwidth as 

\begin{align}\label{repair bandwidth of multiple erasures multiple cosets}
b\leq\frac{d}{n-e}\frac{e\ell}{d-k+e}(n-e+(d-k)(q^a-2)).
\end{align}
Indeed, one can examine the expression of \eqref{average_repair_bandwidth_for_multiple_erasures_in_multiple_cosets}. With the constraint that $\sum_{i=1}^{e} h_i = e$, the first term $h_1(q^a-1)-e)$ is an increasing function of $h_1$ and the second term $(n-h_1(q^a-1))\sum_{i=1}^e \frac{h_i}{d-k+\sum_{v=1}^ih_v}$ is a decreasing function of $h_1$. Under the assumption that $n$ is large, the second term dominates, and increasing $h_1$ reduces the total repair bandwidth $b$. Namely, $h_1=e$ corresponds to the lowest bandwidth for large code length.

In particular, when $d=n-e, h_1=e$, we have
\begin{align}
b=\frac{e\ell}{n-k}(n-e)+\frac{e\ell}{n-k}(n-k-e)(q^a-2),
\end{align}
where the second term is the extra repair bandwidth compared with the MSR bound.

\subsection{Numerical evaluations and discussions}
\label{sec:multiple_numerical}

In this subsection, we compare our schemes for multiple erasures with previous results, including separate repair and schemes in \cite{mardia2018repairing} and \cite{ye2017repairing}.

We first demonstrate that repairing multiple erasures simultaneously can save repair bandwidth compared to repairing erasures separately. Let us assume $e$ failures happen one by one, and the rest of $n-1$ nodes are available as helpers initially when the first failure occurs. We can either repair each failure separately using $n-1$ helpers, or wait for $e$ failures and repair all of them simultaneously with $n-e$ helpers. Table \ref{table of comparision5} shows the comparison. For our scheme in one coset, separate repair needs a repair bandwidth of $\frac{e\ell}{a}(n-1)(a-s)$ symbols in $\mathbb{B}$, simultaneous a repair requires bandwidth of $\frac{e\ell}{a}(n-e)(a-s)$. For our scheme in multiple cosets, we can repair the failures separately by $n-1$ helpers with the bandwidth of $\frac{e\ell}{n-k}[n-1+(n-k-1)(q^a-2)]$, and with simultaneous repair we can achieve the bandwidth of $\frac{e\ell}{n-k}[n-e+(n-k-e)(q^a-2)]$. One can see that in both constructions, simultaneous repair outperforms separate repair.

\begin{table*}
\newcommand{\tabincell}[2]{\begin{tabular}{@{}#1@{}}#2\end{tabular}}
\centering
\caption{\normalfont  Repair bandwidth of different schemes for $e$ erasures. }\label{table of comparision5}
\begin{tabular}{|c|c|c|}

\hline

&repair bandwidth& number of helpers\\
\hline
\tabincell{c} {Single-erasure repair\\ in one coset (separate repair) }& $\frac{e\ell}{a}(n-1)(a-s)$ & $n-1$   \\
\hline
\tabincell{c} {Multiple-erasure repair\\ in one coset (simultaneous repair) }& $\frac{e\ell}{a}(n-e)(a-s)$  & $n-e$  \\
\hline
\tabincell{c} {Single-erasure repair \\ in multiple cosets (separate repair)}& $ \frac{e\ell}{n-k}[n-1+(n-k-1)(q^a-2)]$  & $n-1$\\
\hline
\tabincell{c} {Multiple-erasure repair\\ in multiple cosets (simultaneous repair) }& $\frac{e\ell}{n-k}[n-e+(n-k-e)(q^a-2)]$ & $n-e$  \\
\hline
\end{tabular}
\end{table*}

Nest we compare our scheme for multiple erasures with the existing schemes.  Figure \ref{Comparison of the schemes 3} shows the normalized repair bandwidth for different schemes when $n=16,k=8,e=2,q=2$. Table \ref{table of comparision3} shows the comparison when $n=64,k=32,e=2,q=2$. We make the following observations:

\begin{figure}[!h]
\centering
\includegraphics[width=4in]{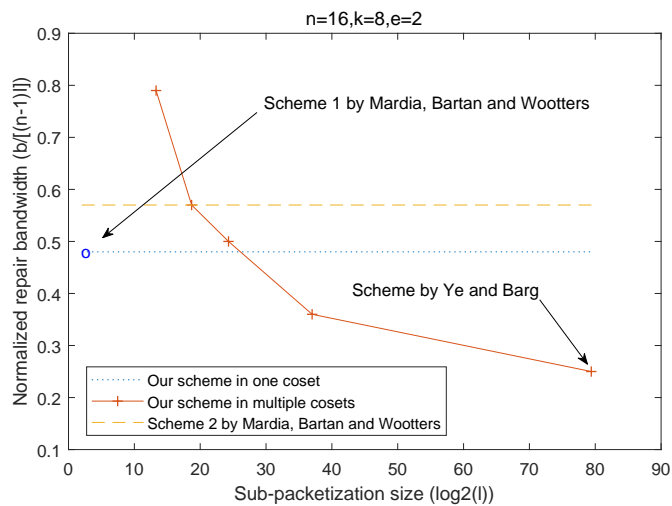}
\caption{Comparison of the schemes, $q=2, n=16,k=8,e=2$, x-axis is the log scale sub-packetization size, y-axis is the normalized repair bandwidth. The scheme by Mardia, Bartan and Wootters is in \cite{mardia2018repairing}, and the scheme by Ye and Barg is in \cite{ye2017repairing}}\label{Comparison of the schemes 3}
\end{figure}

\begin{table*}
\newcommand{\tabincell}[2]{\begin{tabular}{@{}#1@{}}#2\end{tabular}}
\centering
\caption{\normalfont  Normalized repair bandwidth($\frac{b}{(n-e)\ell}$) for different schemes when $n=64,k=32,e=2,q=2$. $\circ$ can be also achieved by Scheme 1 in \cite{mardia2018repairing} and $\ast$ is also achieved by \cite{ye2017repairing}. } \label{table of comparision3}
\begin{tabular}{|c|c|c|c|c|c|c|c|c|}

\hline

& $\ell=6$ & $\ell=7$ &  $\ell=8$ & $\ell=9$ & ...&$\ell=3.6\bl{\times 10}^6$ &$\ell=3.3\times 10^{11}$ & $\ell=3.9\times 10^{115}$ \\
\hline
\tabincell{c} {Normalized bandwidth \\ for Scheme 1 in \cite{mardia2018repairing}   }& 0.42  & 0.50 &0.52  &0.52  &...&0.52&0.52&0.52 \\
\hline
\tabincell{c} {Normalized bandwidth\\  for our scheme  in one coset   }& $0.49^\circ$  & 0.49 &0.49 &0.49 &...&0.49&0.49&0.49\\
\hline
\tabincell{c} {Normalized bandwidth \\ for our scheme  in multiple cosets  }&  &  & &  &  & 0.52&0.48&$0.0625^\ast$\\
\hline

\end{tabular}
\end{table*}

\begin{enumerate}
    \item For fixed $(n,k)$ and our scheme with multiple cosets, we use the paremeter $a$ to adjust the sub-packetization size. From Theorem \ref{theorem for multiple erasures in multiple cosets}, we know that $\ell\approx a\cdot(\frac{n}{q^a-1})^{(\frac{n}{q^a-1})}$, which means that increasing $a$ will decrease the sub-packetization $\ell$. In our schemes with one coset and two cosets, the parameter $a$ is determined by the code length $n$, so increasing $\ell$ will not change $a$ or the normalized repair bandwidth. When $q=2$, our code with $a=1$ coincides with that of \cite{ye2017repairing}.
    \item For small $\ell$ and full-length code ($\ell = \log_2 n$), the scheme in \cite{mardia2018repairing} has the smallest normalized repair bandwidth. (Our scheme in one coset also achieves the same point as Scheme 1 in \cite{mardia2018repairing} when $\ell=\log_2n$.)
    \item When $\ell$ grows larger ($4<\ell < 2.1\times 10^7$ in Figure \ref{Comparison of the schemes 3}, $6 < \ell < 3.3\times 10^{11}$ in Table \ref{table of comparision3}), our scheme in one coset has the smallest repair bandwidth.
    \item For extremely large $\ell$ ($\ell\ge 2.1\times 10^7$ in Figure \ref{Comparison of the schemes 3}, $\ell\ge3.3\times 10^{11}$ in Table \ref{table of comparision3}), our scheme in multiple cosets has the smallest repair bandwidth.
    \item The scheme in \cite{ye2017repairing} also achieves one point in both Figure \ref{Comparison of the schemes 3} and Table \ref{table of comparision3}, which can be viewed as a special case of our scheme in multiple cosets.
\end{enumerate}

\section{Conclusion}
In this paper, we designed three Reed-Solomon code repair schemes to provide a tradeoff between the sub-packetization size and the repair bandwidth. Our schemes choose the evaluation points of the Reed-Solomon code from one, two, or multiple cosets of the multiplicative group of the underlying finite field. For a single erasure, when the sub-packetization size is large, the scheme in multiple cosets has better performance, it approaches the MSR bound. When sub-packetization size is small, the scheme in one coset has advantages in repair bandwidth. The scheme in two cosets has smaller repair bandwidth with certain parameters in between the other two cases. For multiple erasures, our scheme in one coset has constructions for arbitrary redundancy $n-k$ and our scheme in multiple cosets reduced the sub-packetization size of an MSR code. The two schemes together provided a set of tradeoff points and we observe similar tradeoff characteristics as in the single erasure case. In spite of several tradeoff points we provided in this paper, the dependence of the sub-packetization size versus the repair bandwidth is still an open question.

\appendices

\section{Proof of schemes in multiple cosets for the case of arbitrary $a$ and $\ell'$}
In this section, we first introduce a lemma similar to Lemma \ref{ranks for extended fields} that does not require $a$ and $\ell'$ to be relatively prime. By applying this lemma, our constructions in multiple cosets for single and multiple erasures can be generalized when $a$ and $\ell'$ are arbitrary integers. 

We note that a finite field $\mathbb{F}=GF(q^{\ell})$ is also a vector space over $GF(q)$. Let $\mathbb{E}$ be a subspace of $\mathbb{F}$. Define the subspace spanned by a set of elements $\{\gamma_1,\gamma_2,\dots,\gamma_i\} \subseteq \mathbb{F}$ over $\mathbb{E}$ as $\text{span}_\mathbb{E}\{\gamma_1,\gamma_2,\dots,\gamma_i\} \triangleq \{\sum_{j=1}^i b_j\gamma_j: b_j\in \mathbb{E}\}$.
The rank $rank_{\mathbb{E}}(\{\gamma_1,\gamma_2,...,\gamma_i\})$ is defined to be the cardinality of a maximal subset of $\{\gamma_1,\gamma_2,...,\gamma_i\}$ that is linearly independent over $\mathbb{E}$.

\begin{lem}\label{ranks for extended subspace}
Let $\mathbb{B}=GF(q), \mathbb{F}'=GF(q^{\ell'})$, $\mathbb{F}=GF(q^{\ell}), \ell=a \ell'$, and $q$ be any power of a prime number. Define the subspace $\mathbb{E}=\text{span}_\mathbb{B}\{\beta_1,\beta_2,\dots,\beta_a\}$, where $\{\beta_1,\beta_2,\dots,\beta_a\}$ is a basis for $\mathbb{F}$ over $\mathbb{F'}$.
For any set of {$\{\gamma_1,\gamma_2,...,\gamma_{\ell'}\} \subseteq \mathbb{F'}  \le \mathbb{F}$,} we have  
\begin{align}
&rank_{\mathbb{E}}(\{\gamma_1,\gamma_2,...,\gamma_{\ell'}\})\nonumber\\
=&rank_{\mathbb{B}}(\{\gamma_1,\gamma_2,...,\gamma_{\ell'}\}).
\end{align}
\end{lem}

\begin{IEEEproof}
Assume $rank_{\mathbb{B}}(\{\gamma_1,\gamma_2,...,\gamma_{\ell'}\})=c$ and without loss of generality, $\{\gamma_1,\gamma_2,...,\gamma_c\}$ are linearly independent over $\mathbb{B}$. Then, we can construct $\{\gamma'_{c+1},\gamma'_{c+2},...,\gamma'_{\ell'}\}\subseteq\mathbb{F'}$  to make $\{\gamma_1,\gamma_2,...,\gamma_c,\gamma'_{c+1},$  $\gamma'_{c+2},...,\gamma'_{\ell'}\}$ form a basis for $\mathbb{F'}$ over $\mathbb{B}$. 

Since $\{\beta_1,\beta_2,\dots,\beta_a\}$ is the basis for $\mathbb{F}$ over $\mathbb{F'}$, we know that $\{\beta_i\gamma_1,\beta_i\gamma_2,...,\beta_i\gamma_c,\beta_i\gamma'_{c+1},$$\beta_i\gamma'_{c+2},...,$ $\beta_i\gamma'_{\ell'} : i\in[a]\}$ is the basis for $\mathbb{F}$ over $\mathbb{B}$. Then, we have $\mathbb{F}=\text{span}_{\mathbb{E}}\{\gamma_1,\gamma_2,...,\gamma_c,$ $\gamma'_{c+1}, \gamma'_{c+2},...,\gamma'_{\ell'}\}$, namely,
$\{\gamma_1,\gamma_2,...,\gamma_c,\gamma'_{c+1},$  $\gamma'_{c+2},...,\gamma'_{\ell'}\}$ is a basis for $\mathbb{F}$ over $\mathbb{E}$, hence $\{\gamma_1,\gamma_2,...,\gamma_c\}$ are linearly independent over $\mathbb{E}$, 
\begin{align}
&rank_{\mathbb{E}}(\{\gamma_1,\gamma_2,...,\gamma_{\ell'}\})\nonumber\\
\geq&c\nonumber\\
=&rank_{\mathbb{B}}(\{\gamma_1,\gamma_2,...,\gamma_{\ell'}\}).
\end{align}
Since $\mathbb{B}\subseteq\mathbb{E}$, we also have
\begin{align}
&rank_{\mathbb{E}}(\{\gamma_1,\gamma_2,...,\gamma_{\ell'}\})\nonumber\\
\leq&rank_{\mathbb{B}}(\{\gamma_1,\gamma_2,...,\gamma_{\ell'}\}).
\end{align}
The proof is completed. 
\end{IEEEproof}

For the schemes in multiple cosets when $a$ and $\ell'$ are not relatively prime, we just use the subspace $\mathbb{E}=\text{span}_\mathbb{B}\{\beta_1,\beta_2,...,\beta_a\}$ to replace the subfield $GF(q^a)$. We denote by  $\mathbb{E}^\ast=\mathbb{E}\backslash\{0\}$ for the subspace $\mathbb{E}$. The evaluation points of the new code are $\{\gamma\mathbb{E}^\ast: \gamma\in A'\}$ where $A'\subseteq \mathbb{F'}$ is the set of evaluation points for the original code. In the proofs, we use Lemma \ref{ranks for extended subspace} instead of Lemma \ref{ranks for extended fields}.
For example, from Lemma \ref{ranks for extended subspace} we know that the new evaluation points are all distinct if the elements in $A'$ are linearly independent over $\mathbb{B}$.

\section{Proof of Theorem \ref{equivalence}}
\label{sec: appendix equivalence}
In this section, we prove the equivalence of Definitions \ref{linear repair scheme definition} and \ref{dual code repair definition}. We first show that the dual code scheme in Definition \ref{dual code repair definition} reduces to a linear repair scheme as in Definition \ref{linear repair scheme definition} in Lemma \ref{lem: equivalence 4}. Then, we show that Definition \ref{linear repair scheme definition} reduces to  Definition \ref{dual code repair definition} in Lemma \ref{algorithm for multiple erasures proof} and Lemma \ref{lem: equivalence 3}.

\begin{lem} \label{lem: equivalence 4}
The dual code scheme can be reduced to the linear repair scheme in Definition \ref{linear repair scheme definition}.
\end{lem}
\begin{IEEEproof}
In the dual code scheme, we repair nodes $[e]$ from the linearly independent equations
\begin{align}\label{multiple erasures solution}
\sum_{v=1}^e tr_{\mathbb{F}/\mathbb{B}} (C'_{ijv}C_v) = - \sum_{t=e+1}^n tr_{\mathbb{F}/\mathbb{B}} ( C'_{ijt} C_t), i\in[e],j\in[\ell].
\end{align}
Here, $C'_{ijt}$ can be written as
\begin{equation}
C'_{ijt}=\sum\limits_{m=1}^{\ell}\xi'_{m}tr_{\mathbb{F}/\mathbb{B}}(\xi_{m}C'_{ijt}),
\end{equation}
where $\{\xi_{1},\xi_{2},\dots,\xi_{\ell}\}$ and $\{\xi'_{1},\xi'_{2},\dots,\xi'_{\ell}\}$ are the dual basis for $\mathbb{F}$ over $\mathbb{B}$.
Then, we can rewrite \eqref{multiple erasures solution} in matrix form as
\begin{align}\label{matrix dual code}
\sum\limits_{v=1}^{e}S_{iv} \begin{bmatrix}tr_{\mathbb{F}/\mathbb{B}}(\xi'_1C_v)\\tr_{\mathbb{F}/\mathbb{B}}(\xi'_2C_v)\\\vdots\\tr_{\mathbb{F}/\mathbb{B}}(\xi'_\ell C_v)\end{bmatrix}=-\sum\limits_{t=e+1}^{n}S_{it}\begin{bmatrix}tr_{\mathbb{F}/\mathbb{B}}(\xi'_1C_v)\\tr_{\mathbb{F}/\mathbb{B}}(\xi'_2C_v)\\\vdots\\tr_{\mathbb{F}/\mathbb{B}}(\xi'_\ell C_v)\end{bmatrix},i\in[e],
\end{align}
where $S_{it}\in\mathbb{B}^{\ell \times \ell}$ is called the repair matrix defined as
\begin{align}
S_{it}=\begin{bmatrix}tr_{\mathbb{F}/\mathbb{B}}(\xi_{1}C'_{i1t})&\cdots\ &tr_{\mathbb{F}/\mathbb{B}}(\xi_{\ell}C'_{i1t})\\tr_{\mathbb{F}/\mathbb{B}}(\xi_{1}C'_{i2t})&\cdots\ &tr_{\mathbb{F}/\mathbb{B}}(\xi_{\ell}C'_{i2t})\\ \vdots  & \ddots  & \vdots  \\ tr_{\mathbb{F}/\mathbb{B}}(\xi_{1}C'_{i\ell t})&\cdots\ &tr_{\mathbb{F}/\mathbb{B}}(\xi_{\ell}C'_{i\ell t})\end{bmatrix},
\end{align}
Let
\begin{align}
X_w \triangleq \begin{bmatrix}tr_{\mathbb{F}/\mathbb{B}}(\xi'_{1}C_w)\\tr_{\mathbb{F}/\mathbb{B}}(\xi'_{2}C_w)\\\vdots\\tr_{\mathbb{F}/\mathbb{B}}(\xi'_{\ell} C_w)\end{bmatrix}, w\in[n].
\end{align}
We want to solve $X_i, i \in [e]$, which can be used to recover the $e$ failed nodes $C_1,C_2,\dots,C_e$. Define matrix $S$ as 
\begin{align}
S = \begin{bmatrix}S_{11}&S_{12}&\cdots\ &S_{1e}\\S_{21}&S_{22}&\cdots\ &S_{2e}\\ \vdots  & \vdots  & \ddots  & \vdots  \\ S_{e1}&S_{e2}&\cdots\ &S_{ee}\end{bmatrix}.
\end{align}
Then, \eqref{matrix dual code} can be represented as
\begin{align}\label{multiple erasures matrix solution}
    S 
    \begin{bmatrix}
    X_1 \\
    X_2 \\
    \vdots\\
    X_e
    \end{bmatrix}
    = - \sum_{t=e+1}^{n} 
    \begin{bmatrix}S_{1t}\\S_{2t}\\\vdots\\S_{et}\end{bmatrix}
    X_t.
\end{align}
Thus, from Lemma \ref{full rand condition for multiple erasures} we know that if the full rank condition satisfies\footnote{We use Lemma \ref{full rand condition for multiple erasures} while in Lemma \ref{full rand condition for multiple erasures} we use the polynomials $p_{ij}(x)$ as part of the elements in the defined matrix $S_{it}$. However, in Lemma \ref{full rand condition for multiple erasures} we just view the polynomials $p_{ij}(x)$ as a symbol, change them to the dual codeword symbol $C'_{ijt}$ will not have effects on the results of the lemma.}, $S$ has full rank so we can solve $X_i, i \in [e]$. Then, $C_i,i\in[e]$ can be repaired from 
\begin{equation}
C_i=\sum\limits_{m=1}^{\ell}\xi_{m}tr_{\mathbb{F}/\mathbb{B}}(\xi'_{m}C_i),
\end{equation}

Now, set $\mu_{im}=\xi_m$, we get $\lambda_{im}=tr_{\mathbb{F}/\mathbb{B}}(\xi'_{m}C_i)$, which can be solved from \eqref{multiple erasures matrix solution}. Note that the right side of \eqref{multiple erasures matrix solution} is equal to the right side of \eqref{multiple erasures solution}, we get the queries $Q_{t}=\{C'_{ijt},  i\in[e],j\in[\ell]\}$ and the coefficients $\beta_{im\gamma t}$ come from matrix $S$.

Then, we can get the repair bandwidth condition:
\begin{align}
b=\sum_{t=e+1}^n rank_\mathbb{B}(\{C'_{ijt}: i\in[e],j\in[\ell] \})=\sum_{t=e+1}^\ell rank_{\mathbb{B}}(Q_{t}).
\end{align} 
\end{IEEEproof}

\begin{lem}\label{algorithm for multiple erasures proof}
A linear repair scheme in Definition \ref{linear repair scheme definition} can be represented in the form below:
\begin{align}\label{algorithm for multiple erasures}
\mu'_{ij}C_i=\sum_{t=e+1}^n\theta_{ijt}C_t, i \in [e], j\in[\ell],
\end{align}
where $\{\mu'_{i1},\mu'_{i2},...,\mu'_{i\ell}\}$ is the dual basis of $\{\mu_{i1},\mu_{i2},...,\mu_{i\ell}\}$, and $\theta_{ijt}\in \text {span}_\mathbb{B}(Q_{t})$, $e+1 \le t \le n, i \in [e], j\in[\ell]$ are some coefficients in $\mathbb{F}$.
The repair bandwidth is 
\begin{align} \label{eq: bandwidth 110}
b = \sum_{t=e+1}^n rank_{\mathbb{B}} (\{\theta_{ijt}: i \in [e], j\in[\ell] \}).
\end{align}

\end{lem}
\begin{IEEEproof}
By \eqref{eq:66} and \eqref{eq:67} we have
\begin{align}\label{eq:110}
\sum_{t=e+1}^n \sum_{\gamma\in Q_{t}} tr_{\mathbb{F}/\mathbb{B}}( \beta_{ij\gamma t} \cdot \gamma C_t)=\lambda_{ij}
=tr_{\mathbb{F}/\mathbb{B}}(\mu'_{ij}C_i).
\end{align}
Set $\theta_{ijt}=\sum_{\gamma\in Q_{t}} \beta_{ij\gamma t} \cdot \gamma$. Then, we have $\theta_{ijt}\in \text {span}_\mathbb{B}(Q_{t})$.
Hence,
\begin{align}\label{eta choose equation}
\sum_{t=e+1}^ntr_{\mathbb{F}/\mathbb{B}}(\theta_{ijt}C_t)=\sum_{t=e+1}^n  tr_{\mathbb{F}/\mathbb{B}}(\sum_{\gamma\in Q_{t}} \beta_{ij\gamma t} \cdot \gamma C_t)=tr_{\mathbb{F}/\mathbb{B}}(\mu'_{ij}C_i).
\end{align}

Equations \eqref{eq:110} and \eqref{eta choose equation} hold for all $f\in \mathbb{F}[x]$. Since the RS code is a linear code, they also hold for $\delta_m \cdot f\in \mathbb{F}[x]$ for all $\delta_m \in \mathbb{F}$. In particular, let $\delta_m, m \in [\ell],$ be a basis for $\mathbb{F}$ over $\mathbb{B}$, Then,
\begin{align}
tr_{\mathbb{F}/\mathbb{B}}(\delta_m \cdot\mu'_{ij}C_i)=tr_{\mathbb{F}/\mathbb{B}}(\delta_m \cdot\sum_{t=e+1}^n\theta_{ijt}C_t), \forall m \in [\ell],
\end{align}
which in turn implies that $\theta_{ijt}$ also satisfies \eqref{algorithm for multiple erasures}.
\end{IEEEproof}

Note that the repair bandwidth \eqref{eq: bandwidth 110} also satisfies
\begin{align}\label{eq: bandwidth inequality}
b &= \sum_{t=e+1}^n rank_{\mathbb{B}} (\theta_{ijt}: i \in [e], j\in[\ell]) \nonumber \\
&\le  \sum_{t=e+1}^n rank_{\mathbb{B}} (Q_{t}), 
\end{align}
since $\theta_{ijt} \in \text {span}_\mathbb{B}(Q_{t})$. However, for any linear scheme $L$ in Definition \ref{linear repair scheme definition}, if \eqref{eq: bandwidth inequality} holds with strict inequality, we can improve the linear scheme $L$ by setting $Q_{t}$ such that $\text {span}_\mathbb{B}(Q_{t}) = \text {span}_\mathbb{B} (\{\theta_{ijt}, i \in [e], j\in[\ell]\}),$ for all $e+1 \le t \le n$. Hence, the linear scheme $L$ and the scheme in Lemma \ref{algorithm for multiple erasures proof} have identical bandwidth.

\begin{lem} \label{lem: equivalence 3}
The scheme in Lemma \ref{algorithm for multiple erasures proof} can be represented by the dual code scheme in Definition \ref{dual code repair definition}.
\end{lem}
\begin{IEEEproof}
By \eqref{algorithm for multiple erasures}, $(0,\dots,\mu'_{ij},\dots,0, \theta_{ije+1},\dots,\theta_{ijn})$ is a dual codeword, where $\mu'_{ij}$ is the $i$-th entry. Then, for $j\in[\ell]$, we set $C'_{ijt}$ such that $C'_{iji}=-\mu'_{ij}$, $C'_{ijv}=0, v \in [e],v\neq i$ and $C'_{ijt}= \theta_{ijt}, e+1 \le t \le n$. The full rank condition follows because $\{\mu'_{i1},\mu'_{i2},...,\mu'_{i\ell}\}$ is the basis for $\mathbb{F}$ over $\mathbb{B}$, and the repair bandwidth condition follows from \eqref{eq: bandwidth inequality}.
Thus, we obtain the dual code scheme in Definition \ref{dual code repair definition}.

\end{IEEEproof}

\section{Proof of Lemma \ref{full rand condition for multiple erasures}}

\begin{IEEEproof}
Vectors $V_{ij},i\in[e],j\in[\ell]$ are linearly independent over $\mathbb{B}$ is equivalent to that there is no nonzero $b_{ij}\in\mathbb{B},i\in[e],j\in[\ell]$ that satisfy
\begin{align}
\sum_{i,j}b_{ij}p_{ij}(\alpha_v)=0, \forall v\in[e].
\end{align}
Here, $p_{ij}(x)$ can be written as
\begin{equation}\label{dual basis equation2}
p_{ij}(x)=\sum\limits_{m=1}^{\ell}\xi'_{m}tr_{\mathbb{F}/\mathbb{B}}(\xi_{m}p_{ij}(x)),
\end{equation}
where $\{\xi_{1},\xi_{2},\dots,\xi_{\ell}\}$ and $\{\xi'_{1},\xi'_{2},\dots,\xi'_{\ell}\}$ are the dual basis for $\mathbb{F}$ over $\mathbb{B}$. So, it is equivalent to that there is no nonzero $b_{ij}\in\mathbb{B},i\in[e],j\in[\ell]$ that satisfy
\begin{align}
\sum_{i,j}b_{ij}\sum\limits_{m=1}^{\ell}\xi'_{m}tr_{\mathbb{F}/\mathbb{B}}(\xi_{m}p_{ij}(\alpha_v))=0, \forall v\in[e].
\end{align}
Since $\{\xi'_{1},\xi'_{2},\dots,\xi'_{\ell}\}$ are linearly independent over $\mathbb{B}$. Therefore, there is no nonzero $b_{ij}\in\mathbb{B},i\in[e],j\in[\ell]$ that satisfy
\begin{align}
\sum_{i,j}b_{ij}tr_{\mathbb{F}/\mathbb{B}}(\xi_{m}p_{ij}(\alpha_v))=0, \forall v\in[e],m\in[\ell],
\end{align}
which is equivalent to $S$ has full rank.
\end{IEEEproof}

\section{Proof of Lemma \ref{matrix rank}}

\begin{IEEEproof}
Assume $rank_\mathbb{B}(\{p_{ij}(\alpha_t),i\in[e],j\in[\ell]\})=c$ and $\{p_{ij}(\alpha_t),(i,j)\in I\}$ are linearly independent over $\mathbb{B}, |I|=c$. Define $S_{it}(j)$ as the vector for the $j$-th row in $S_{it}$: $S_{it}(j)=(tr_{\mathbb{F}/\mathbb{B}}(\xi_{1}p_{ij}(\alpha_t))$, $tr_{\mathbb{F}/\mathbb{B}}(\xi_{2}p_{ij}(\alpha_t)),...,tr_{\mathbb{F}/\mathbb{B}}(\xi_{\ell}p_{ij}(\alpha_t)))$. We first prove $\{S_{it}(j),(i,j)\in I\}$ are linearly independent and then prove $S_{i't}(j'),i'\in[e],j'\in[\ell],(i',j')\notin I$ can be represented as $\mathbb{B}$-linear combinations of $\{S_{it}(j),(i,j)\in I\}$ .

If $\{S_{it}(j),(i,j)\in I\}$ are linearly dependent over $\mathbb{B}$, then there exists some nonzero $b_{ij}\in\mathbb{B},(i,j)\in I$ that satisfies
\begin{align}
\sum_{(i,j)\in I} b_{ij}S_{it}(j)=0,
\end{align}
and we have
\begin{align}
\sum_{(i,j)\in I} b_{ij}tr_{\mathbb{F}/\mathbb{B}}(\xi_{m}p_{ij}(\alpha_t))=0, \forall m\in[\ell].
\end{align}
Multiplying the above equation by $\xi'_{m}$ and summing over all $m\in[\ell]$ result in
\begin{align}
\sum_{(i,j)\in I} \sum_{m=1}^\ell b_{ij}\xi'_{m}tr_{\mathbb{F}/\mathbb{B}}(\xi_{m}p_{ij}(\alpha_t))=0.
\end{align}
Then, from \eqref{dual basis equation2} we know that $b_{ij}$ satisfies
\begin{align}
\sum_{(i,j)\in I} b_{ij}p_{ij}(\alpha_t)=0,
\end{align}
which is contradictory to the statement that $\{p_{ij}(\alpha_t),(i,j)\in I\}$ are linearly independent over $\mathbb{B}$. Therefore, 
$\{S_{it}(j),(i,j)\in I\}$
are linearly independent over $\mathbb{B}$.

Let us assume $p_{i'j'}(\alpha_t),i'\in[e],j'\in[\ell],(i',j')\notin I$ can be represented as
\begin{align}
p_{i'j'}(\alpha_t)=\sum_{(i,j)\in I} b'_{ij}p_{ij}(\alpha_t), \text{ for some } b'_{ij}\in\mathbb{B}.
\end{align}
Then, for $m\in[\ell]$,
\begin{align}
tr_{\mathbb{F}/\mathbb{B}}(\xi_{m}p_{i'j'}(\alpha_t))=tr_{\mathbb{F}/\mathbb{B}}\left(\xi_{m}\sum_{(i,j)\in I} b'_{ij}p_{ij}(\alpha_t)\right)\nonumber\\
=\sum_{(i,j)\in I} b'_{ij}tr_{\mathbb{F}/\mathbb{B}}(\xi_{m}p_{ij}(\alpha_t)),
\end{align}
which means that for $i'\in[e],j'\in[\ell],(i',j')\notin I$,
\begin{align}
S_{i't}(j')=\sum_{(i,j)\in I} b'_{ij}S_{it}(j)
\end{align}
is the $\mathbb{B}$-linear combination of $\{S_{it}(j),(i,j)\in I\}$.
\end{IEEEproof}

\bibliographystyle{IEEEtran}
\bibliography{IEEEabrv,sample}

\end{document}